\documentclass[pre,showpacs,showkeys,preprintnumbers,amsmath,amssymb,superscriptaddress,twocolumn]{revtex4-2}
\pdfoutput=1 %This is for the ArXiv submission only
\usepackage{graphicx}
\usepackage{amsfonts}
\usepackage{url}
\usepackage{epstopdf}
\usepackage{color}
\usepackage{tikz-cd}
\usepackage{bm}
\usepackage[colorlinks=true,linkcolor=blue,citecolor=red]{hyperref}

\newcommand{\clb}{\color{blue}}

\def\Emax{{\mathcal E}_{\rm max}}

\def\E{{\mathbb E}}
\def\P{{\mathbb P}}
\def\R{{\mathbb R}}

\def\GG{{\bf G}}
\def\MM{{\bf M}}
\def\VV{{\bf V}}

\def\O{\mathcal{O}}

\def\n{\bm{n}}
\def\ttau{\bm{\tau}}
\def\x{\bm{x}}
\def\y{\bm{y}}
\def\v{\bm{v}}

\def\G{\mathcal{G}}

\def\X{\bm{X}}
\def\W{\bm{W}}

\def\A{\mathcal{A}}

\begin{document}

\title{Reactive capacitance of flat patches of arbitrary shape}

\author{Denis S. Grebenkov}
 \email{denis.grebenkov@polytechnique.edu}
\affiliation{
Laboratoire de Physique de la Mati\`{e}re Condens\'{e}e (UMR 7643), \\ 
CNRS -- Ecole Polytechnique, Institut Polytechnique de Paris, 91120 Palaiseau, France}

\author{Raphael Maurette}
 \email{raphael.maurette.public@gmail.com}
\affiliation{
Laboratoire de Physique de la Mati\`{e}re Condens\'{e}e (UMR 7643), \\ 
CNRS -- Ecole Polytechnique, Institut Polytechnique de Paris, 91120 Palaiseau, France}

\date{\today}

\begin{abstract}
We investigate the capacity of a flat partially reactive patch of
arbitrary shape to trap independent particles that undergo
steady-state diffusion in the three-dimensional space.  We focus on
the total flux of particles onto the patch that determines its
reactive capacitance.  To disentangle the respective roles of the
reactivity and the shape of the patch, we employ a spectral expansion
of the reactive capacitance over a suitable Steklov eigenvalue
problem.  We derive several bounds on the reactive capacitance to
reveal its monotonicity with respect to the reactivity and the shape.
Two probabilistic interpretations are presented as well.  An efficient
numerical tool is developed for solving the associated Steklov
spectral problem for patches of arbitrary shape.  We propose and
validate, both theoretically and numerically, a simple, fully explicit
approximation for the reactive capacitance that depends only on the
surface area and the electrostatic capacitance of the patch.  This
approximation opens promising ways to access various characteristics
of diffusion-controlled reactions in general domains with multiple
small well-separated patches.  Direct applications of these results in
statistical physics and physical chemistry are discussed.
\end{abstract}

\pacs{02.50.-r, 05.40.-a, 02.70.Rr, 05.10.Gg}

%02.50.-r       (Probability theory, stochastic processes, and statistics)
%05.40.-a 	Fluctuation phenomena, random processes, noise, and Brownian motion
%02.70.Rr       (General statistical methods)
%05.10.Gg 	Stochastic analysis methods (Fokker-Planck, Langevin, etc.) 

%02.50.Ey 	Stochastic processes  (Probability theory, stochastic processes, and statistics)

\keywords{diffusion, capacitance, Steklov problem, partial reactivity, Robin boundary condition, first-reaction time}

\maketitle

\section{Introduction}

In many biochemical scenarios, a particle diffuses in a complex medium
and searches for small reactive patches that are dispersed on
reflecting walls \cite{House,Murrey,Schuss,Bressloff13,Lindenberg}.
Depending on the considered application, the reactive patch can
represent an ion channel, an active site or a catalytic germ, a hole,
a magnetic impurity, etc.  The mean first-passage time (MFPT) to the
patch is often used to characterize the efficiency of the diffusive
search \cite{Redner,Metzler,Masoliver,Grebenkov,Dagdug}.  Starting
from the Lord Rayleigh's seminal result for the MFPT to a small
circular patch on the sphere \cite{Rayleigh}, considerable progress
has been achieved in solving this so-called narrow escape problem
\cite{Grigoriev02,Holcman04,Singer06a,Singer06b,Singer06c,Schuss07,Benichou08,Singer08,Reingruber09,Caginalp12}
(see also overviews in \cite{Holcman13,Holcman14}).  Among various
analytical tools, one can mention the method of matched asymptotic
expansions
\cite{Pillay10,Cheviakov10,Cheviakov12,Gomez15,Lindsay17,Lindsay18,Bernoff18b,Bressloff22},
homogenization techniques
\cite{Berg77,Berezhkovskii04,Berezhkovskii06,Muratov08,Bernoff18,Plunkett24},
constant-flux approximation
\cite{Shoup81,Grebenkov17,Grebenkov17b,Grebenkov18,Grebenkov19,Grebenkov21},
Wiener-Hopf integral equation \cite{Guerin23}, as well as conformal
mapping in two dimensions \cite{Grebenkov16,Marshall16}.  Other
characteristics of diffusion-controlled reactions
\cite{North66,Wilemski73,Calef83,Berg85,Rice85,Grebenkov23c} such as
the overall reaction rate (or the total flux) and splitting
probabilities were also analyzed.

Former works were mainly focused on the idealized situation of
perfectly reactive patches when the reaction event occurs immediately
upon the first contact with the patch.  In other words, the kinetic
step of the reaction event was ignored, by assuming that it occurs
much faster than the diffusion step.  However, in most practical
settings, once the particle has arrived onto the patch, it has to
overpass an energy activation barrier to react, or an entropic barrier
to squeeze through a hole.  In biochemistry, a macromolecule has to be
in the proper conformational state to bind its partner, an ion channel
has to be open to pass the arrived ion, and so on.  As discussed in
\cite{Grebenkov17}, the ignorance of this kinetic step may cause
misleading conclusions on the asymptotic behavior of the mean
first-reaction time (MFRT) and other characteristics.  The important
role of partial reactivity in diffusion-controlled reactions, first
recognized in the seminal work by Collins and Kimball
\cite{Collins49}, has been revealed in different contexts
\cite{Sano79,Sano81,Shoup82,Zwanzig90,Bressloff08,Singer08b,Lawley15,Piazza19,Grebenkov20,Guerin21,Piazza22}.
In particular, the asymptotic behavior of the MFRT on partially
reactive targets was shown to be considerably different from the
idealized case of perfect reactions
\cite{Grebenkov20,Cengiz24,Grebenkov25b}.
Recently, the problem of diffusion-controlled reactions on multiple
small well-separated partially reactive patches in both two and three
dimensions was revisited \cite{GrebWard25a,GrebWard25c,GrebWard25b}.
In particular, the MFRT in three dimensions was shown to be
determined, to the leading order, by the {\it reactive capacitance},
which is proportional to the flux of particles reacted on the patch
and thus naturally generalizes the electrostatic capacitance to the
case of partial reactivity (see Sec. \ref{sec:theory} for details).
For a circular patch, the reactive capacitance was thoroughly
investigated in \cite{GrebWard25a}.

In this paper, we push this analysis to another direction and
investigate the reactive capacitance for patches of various shapes.
First, we deepen the theoretical description of the reactive
capacitance in Sec. \ref{sec:theory} by uncovering its monotonicity
properties and probabilistic interpretations.  We also get useful
bounds on the reactive capacitance and the principal Steklov
eigenvalue.  For this purpose, we reformulate the exterior Steklov
problem for flat patches as an eigenvalue problem for an integral
operator with an explicit kernel.  Second, we employ this
reformulation to conceive an efficient method for solving the exterior
Steklov problem numerically.  Various numerical results are present in
Sec. \ref{sec:numerics}.  In particular, we inspect how the Steklov
eigenvalues and eigenfunctions depend on the shape of the patch by
considering several families of shapes such as ellipses and
rectangles.  We show that the principal eigenfunction provides the
dominant contribution to the reactive capacitance and validate a
simple, fully explicit approximation for this quantity.  Section
\ref{sec:conclusion} discusses physical implications of this analysis
and concludes the paper by summarizing the main results and
highlighting open questions.

\begin{figure}
\begin{center}
\includegraphics[width=0.58\columnwidth]{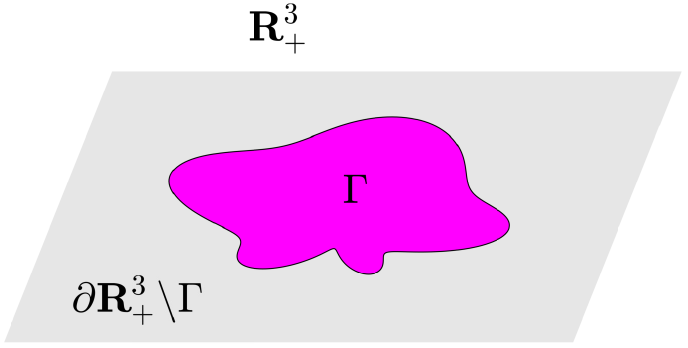} % {scheme2new2.eps}
\end{center}
\caption{
Steady-state diffusion in the upper half-space $\R^3_+$ towards a flat
reactive patch $\Gamma$ on the horizontal reflecting plane.}
\label{fig:scheme}
% A_Raphael_DN_scheme1;
% A_Raphael_DN_scheme2;
\end{figure}

\section{Reactive capacitance}
\label{sec:theory}

Throughout this paper, we consider a flat partially reactive bounded
patch $\Gamma$ of arbitrary shape on the horizontal reflecting plane
in the upper half-space $\R^3_+$ (Fig. \ref{fig:scheme}).  To avoid
eventual mathematical issues on the well-posedness of the considered
problems, we assume that the boundary of the patch is piecewise
smooth.  We consider steady-state diffusion of particles with constant
diffusion coefficient $D > 0$, concentration $A_0 > 0$ imposed at
infinity, and reactivity $\kappa > 0$ on the patch.  The concentration
profile in the upper half-space can be expressed as $A(\y) = A_0(1 -
w(\y;\mu))$, where $\mu = \kappa/D$ and the function $w(\y;\mu)$
solves the boundary value problem
\begin{subequations}  \label{eq:wi_def}
\begin{align}  \label{eq:wi_eq}
\Delta w & = 0 \quad (\y \in \R^3_+), \\  \label{eq:wi_Robin}
\partial_n w + \mu w & = \mu  \quad (\y \in \Gamma), \\  \label{eq:wi_Neumann}
\partial_n w & = 0 \quad (\y \in \partial \R^3_+ \backslash \overline{\Gamma}), \\  \label{eq:wi_inf}
w & \sim \frac{C(\mu)}{|\y|} + \O(|\y|^{-2}) \quad (|\y|\to\infty),
\end{align}
\end{subequations}
where $\y = (y_1,y_2,y_3)$ and $\partial_n = -\partial_{y_3}$ is the
normal derivative.  The coefficient $C(\mu)$ in the far-field
behavior of the solution is called the reactive capacitance of the
patch $\Gamma$.  More explicitly, the divergence theorem yields
\begin{equation}  \label{eq:Ci_def}
C(\mu) = \frac{1}{2\pi} \int\limits_{\Gamma} (\partial_n w) d\y ,
\end{equation}
i.e., $C(\mu)$ is proportional to the flux of particles onto a
partially reactive patch $\Gamma$.  In the limit $\mu\to \infty$, one
retrieves the electrostatic capacitance (also known as harmonic or
Newtonian capacity) of the patch $\Gamma$ \cite{Landkof}, as if it was
perfectly reactive (absorbing).  Note that the solution $w$ can be
extended to the lower half-space by symmetry so that the considered
setting is actually equivalent to a ``two-sided'' patch in the whole
space, without the auxiliary Neumann condition (\ref{eq:wi_Neumann}).
However, as the patch is flat (of zero thickness), it is more
convenient to deal with the problem (\ref{eq:wi_def}) in the upper
half-space.

Even for a circular patch, there is no explicit solution of the
boundary value problem (\ref{eq:wi_def}).  In order to investigate the
dependence of the reactive capacitance on the reactivity parameter
$\mu$ and the patch shape, we will employ spectral expansions,
variational formulations, probabilistic interpretations and numerical
analysis.

\subsection{Exterior Steklov problem}
\label{sec:Steklov}

In order to obtain a spectral expansion of the reactive capacitance,
we consider the exterior Steklov problem in the upper half-space
$\R^3_+$, which consists in finding the eigenpairs $\{\mu_k, \Psi_k\}$
that satisfy \cite{GrebWard25a}
\begin{subequations}  \label{eq:Psi_def}
\begin{align}  \label{eq:Psi_Laplace}
\Delta \Psi_k & = 0 \quad (\y \in \R^3_+), \\
\partial_n \Psi_k & = \mu_k \Psi_k  \quad (\y \in \Gamma), \\  \label{eq:Psi_Neumann}
\partial_n \Psi_k & = 0 \quad (\y \in \partial \R^3_+ \backslash \overline{\Gamma}), \\  \label{eq:Psi_inf}
\Psi_k & \to 0 \quad (|\y|\to\infty).
\end{align}
\end{subequations}
This is a special case of the Steklov eigenvalue problem for the
exterior of a compact domain \cite{Auchmuty14,Arendt15,Bundrock25},
which is known to have a discrete spectrum, i.e., a countable set of
eigenvalues $\mu_k$ that can be enumerated by an integer index $k =
0,1,\ldots$ in an increasing order:
\begin{equation}
0 < \mu_0 \leq \mu_1 \leq \ldots \nearrow +\infty.
\end{equation}
The first eigenvalue $\mu_0$ is simple (of multiplicity $1$) and
strictly positive, whereas the associated eigenfunction $\Psi_0$ does
not change sign.  The restrictions of the Steklov eigenfunctions onto
the patch $\Gamma$, $\Psi_k|_{\Gamma}$, form a complete orthonormal
basis of the $L^2(\Gamma)$ space of square-integrable functions on
$\Gamma$:
\begin{equation}
\int\limits_{\Gamma} \Psi_k(\y) \, \Psi_{k'}(\y) d\y = \delta_{k,k'} 
\end{equation}
(see \cite{Levitin,Girouard17,Colbois24} for a broad overview of
spectral properties of the conventional Steklov problem for bounded
domains).

To gain a more practical insight onto this spectral problem, let us
introduce the Green's function $G(\y,\y')$ in the upper half-space
$\R^3_+$ with Neumann boundary condition on the horizontal plane
$\partial \R^3_+$:
\begin{subequations}
\begin{align}  \label{eq:Green}
-\Delta G(\y,\y') & = \delta(\y-\y') \quad (\y\in\R^3_+), \\
\partial_n G(\y,\y') & = 0 \quad (\y\in \partial\R^3_+), \\
G(\y,\y') & \to 0 \quad (|\y|\to \infty),
\end{align}
\end{subequations}
where $\delta(\y-\y')$ is the Dirac distribution.  By the method of
images, one gets immediately
\begin{equation}  \label{eq:Green_half}
G(\y,\y') = \frac{1}{4\pi} \biggl(\frac{1}{|\y-\y'|} + \frac{1}{|\y - \bar{\y}'|}\biggr),
\end{equation}
where $\bar{\y}' = (y'_1,y'_2,-y'_3)$ is the mirror reflection of $\y'
= (y'_1,y'_2,y'_3)$ with respect to the horizontal plane.  Multiplying
Eq. (\ref{eq:Psi_Laplace}) by $G(\y,\y')$, multiplying
Eq. (\ref{eq:Green}) by $\Psi_k(\y)$, subtracting these equations,
integrating them over $\y\in\R^3_+$, using the Green's formula and the
boundary conditions, one finds
\begin{equation}  \label{eq:Psi_G}
\Psi_k(\y') = \mu_k \int\limits_{\Gamma} G(\y,\y') \Psi_k(\y) d\y  \quad (\y'\in \overline{\R^3_+}).
\end{equation} 
Restricting $\y'$ to $\Gamma$, one can introduce an integral operator
$\G$, acting on a continuous function $f(\y)$ on $\Gamma$ as
\begin{equation}
(\G f)(\y) = \int\limits_{\Gamma} \G(\y,\y') f(\y') d\y' ,
\end{equation}
with the kernel 
\begin{equation}  \label{eq:G_kernel}
\G(\y,\y') = \frac{1}{2\pi |\y-\y'|}  \qquad (\y, \y'\in \Gamma, ~ \y \ne \y')
\end{equation}
(even though the operator $\G$ is formally defined on the space of
continuous functions, it can be extended to $L^2(\Gamma)$ space via
embedding theorems, see \cite{Kress} for details).
Since the kernel $\G(\y,\y')$ is symmetric and weakly singular, $\G$
is a self-adjoint compact operator \cite{Kress};
% see Th. 2.22, page 24
moreover, if $\Gamma$ was replaced by the whole plane $\R^2$, $\G$
would be the Riesz potential of order $1$ \cite{Adams}, which is known
to be positive definite (this is a direct corollary of the Fourier
transform representation of Riesz potentials, see \cite{Stein}).  As a
consequence, there is an infinite sequence of positive eigenvalues,
enumerated by the index $k = 0,1,2,\ldots$ that accumulate to $0$,
whereas the associated eigenfunctions form an orthonormal basis of
$L^2(\Gamma)$.  Rewriting Eq. (\ref{eq:Psi_G}) as
\begin{equation}  \label{eq:Psi_integral}
\G \Psi_k|_{\Gamma} = \frac{1}{\mu_k} \Psi_k|_{\Gamma}  \qquad (k=0,1,2,\ldots),
\end{equation} 
one sees that $1/\mu_k$ and $\Psi_k|_{\Gamma}$ are the eigenvalues and
eigenfunctions of $\G$.  This is an alternative way to access the
eigenvalues and eigenfunctions of the Steklov problem
(\ref{eq:Psi_def}); note that once $\Psi_k|_{\Gamma}$ is found, its
extension to the upper half-space follows immediately from
Eq. (\ref{eq:Psi_G}).  We finally stress that the integral operator
$\G$ is the inverse of the Dirichlet-to-Neumann operator associated to
the exterior Steklov problem (\ref{eq:Psi_def}).

This reformulation reduces the original three-dimensional problem in
the unbounded domain (the upper half-space) to an equivalent
eigenvalue problem for the integral operator $\G$ with the explicit
kernel on the two-dimensional bounded patch $\Gamma$.  Note that a
modified integral equation with the kernel (\ref{eq:G_kernel}), in
which the right-hand side is a constant, was employed by Bernoff and
Lindsay to investigate diffusive capture rates for an array of
perfectly reactive circular patches \cite{Bernoff18b}.  The eigenvalue
problem (\ref{eq:Psi_integral}) will serve us to develop an efficient
numerical tool for computing the Steklov eigenfunctions (see Appendix
\ref{sec:FEM}).  Moreover, it yields a variational (minimax)
characterization of the Steklov eigenvalues \cite{Kress}
% p. 276, Th. 15.14 (Courant)
%
\begin{subequations}
\begin{align}
\frac{1}{\mu_0} & = \sup\limits_{\|u\| = 1}  \bigl\{ (\G u,u)_{L^2(\Gamma)} \bigr\} ,\\  \label{eq:muk_minimax}
\frac{1}{\mu_k} & = \inf\limits_{u_1,\ldots,u_k} \sup\limits_{u\perp \{u_1,\ldots,u_k\} \atop \|u\| = 1} 
 \bigl\{ (\G u,u)_{L^2(\Gamma)} \bigr\}  \quad (k = 1,2\ldots),
\end{align} 
\end{subequations}
for test functions $u$ and $u_k$ from $L^2(\Gamma)$ of the unit
$L^2(\Gamma)$-norm.  This variational form implies the domain
monotonicity property: 
\begin{equation}  \label{eq:muk_monotonicity}
\Gamma_1 \subset \Gamma_2  \quad \Rightarrow \quad \mu_k^{\Gamma_1} \geq \mu_k^{\Gamma_2}  \quad (k=0,1,\ldots).
\end{equation}
This property can be shown by a standard construction, namely, by
using a test function $u(\y)$, which is equal to the $k$-th
eigenfunction $\Psi_k^{\Gamma_1}(\y)$ for $\y\in \Gamma_1$ and $0$
otherwise:
\begin{equation*}
\frac{1}{\mu_k^{\Gamma_1}} = \bigl(\G \Psi_k^{\Gamma_1}|_{\Gamma_1},\Psi_k^{\Gamma_1}|_{\Gamma_1}\bigr)_{L^2(\Gamma_1)}
= \bigl(\G u, u\bigr)_{L^2(\Gamma_2)} \leq \frac{1}{\mu_k^{\Gamma_2}} \,,
\end{equation*}
where the last inequality follows from the minimax characterization
(\ref{eq:muk_minimax}).  

We emphasize that the domain monotonicity property does not hold for
the Steklov problem in a more general setting of the exterior of a
compact set.  For instance, if $\Gamma$ is an oblate spheroidal
surface with semi-axes $a,a,b$, the domain monotonicity property
(\ref{eq:muk_monotonicity}) fails, as illustrated in Fig. 7 of
\cite{Grebenkov24}, which shows the non-monotonous dependence of the
principal Steklov eigenvalue on the aspect ratio $a/b$.  This example
highlights that the domain monotonicity property holds for patches due
to their flatness.

\subsection{Spectral expansion and basic properties}
\label{sec:reactive}

Since the Steklov eigenfunctions $\Psi_k|_\Gamma$ form the orthonormal
basis of $L^2(\Gamma)$, they can be used to expand
$w(\y;\mu)|_{\Gamma}$ and thus to determine the normal derivative of
$w(\y;\mu)$.  Substituting this expansion into Eq. (\ref{eq:Ci_def}),
one gets the spectral representation of the reactive capacitance
(see \cite{GrebWard25a,Grebenkov25a} for details)
\begin{equation}  \label{eq:Cmu}
C(\mu) = \frac{\mu |\Gamma|}{2\pi} \sum\limits_{k=0}^\infty \frac{\mu_k F_k}{\mu_k + \mu} \,,  \quad
F_k = \frac{1}{|\Gamma|} \biggl(\int\limits_{\Gamma} \Psi_k(\y) d\y\biggr)^2.
\end{equation}
By expanding a constant function on the basis of eigenfunctions
$\Psi_k|_{\Gamma}$, one can easily check that
\begin{equation}  \label{eq:Fk_sum}
\sum\limits_{k=0}^\infty F_k = 1,
\end{equation}
so that the coefficient $F_k \geq 0$ of the spectral expansion
(\ref{eq:Cmu}) can be interpreted as the relative contribution of the
$k$-th Steklov eigenfunction to the reactive capacitance $C(\mu)$.  In
particular, the $F_0$ characterizes the closeness of the principal
eigenfunction $\Psi_0$ to a constant.  A similar spectral expansion
was earlier derived for the spectroscopic impedance of an
electrochemical cell, though in a slightly different mathematical
setting \cite{Grebenkov06b}.

Evaluating the derivative of $C(\mu)$,
\begin{equation}
\frac{dC(\mu)}{d\mu} = \frac{|\Gamma|}{2\pi} \sum\limits_{k=0}^\infty \frac{\mu_k^2 F_k}{(\mu_k+\mu)^2} > 0 ,
\end{equation}
one sees that $C(\mu)$ is a monotonously growing function of $\mu$ for
all $\mu \geq 0$.  A Taylor expansion of $C(\mu)$ around $0$ reads
\begin{equation}  \label{eq:Cmu_Taylor}
C(\mu) = - \sum\limits_{n=1}^\infty c_n (-\mu)^n , \quad \textrm{with} ~ c_n = \frac{|\Gamma|}{2\pi} 
\sum\limits_{k=0}^\infty \frac{F_k}{\mu_k^{n-1}} \,.
\end{equation}
In \cite{GrebWard25a}, alternative representations for the first three
coefficients were derived:
\begin{equation}  \label{eq:cn_int}
c_1 = \frac{|\Gamma|}{2\pi}, ~~
c_2 = \frac{1}{2\pi} \int\limits_{\Gamma} \omega_\Gamma(\y) \, d\y, 
~~
c_3 = \frac{1}{2\pi} \int\limits_{\Gamma} \omega_\Gamma^2(\y) \, d\y, 
\end{equation}
where
\begin{equation}  \label{eq:w_def}
\omega_\Gamma(\y) = \int\limits_{\Gamma} \frac{d\y'}{2\pi |\y-\y'|} \,.
\end{equation}
In the following, we will also employ the notation:
\begin{equation}  \label{eq:A_def}
\A_\Gamma = \frac{1}{|\Gamma|^2} \int\limits_{\Gamma} \omega_\Gamma(\y) d\y
= \frac{1}{|\Gamma|^2} \int\limits_{\Gamma \times\Gamma} \G(\y,\y') d\y d\y' .
\end{equation}
This constant, re-emerging in various contexts \cite{Grebenkov25b},
will play an important role in the analysis of the reactive
capacitance; note that $c_2 = \A_\Gamma |\Gamma|^2/(2\pi)$.

In the opposite limit of large $\mu$, Eq. (\ref{eq:Cmu}) yields the
spectral expansion for the electrostatic capacitance:
\begin{equation}  \label{eq:Fkmuk_sum}
C(\infty) = \frac{|\Gamma|}{2\pi} \sum\limits_{k=0}^\infty \mu_k F_k 
\end{equation}
(here we employ the convention that the electrostatic capacitance of a
ball of radius $R$ is equal to $R$).

If the patch is dilated by a factor $\alpha$, i.e., $\Gamma' = \alpha
\Gamma$, the Steklov eigenvalues are rescaled by $1/\alpha$, $\mu_k' =
\mu_k/\alpha$, $F_k$ remain unchanged, so that the reactive
capacitance of the rescaled patch reads
\begin{equation}  \label{eq:Cmu_scaling}
C'(\mu) = \alpha \, C(\alpha \mu).
\end{equation}

\subsection{Alternative spectral expansion}
\label{sec:Neumann}

The reactive capacitance $C(\mu)$ admits an alternative spectral
expansion, which is based on the eigenpairs $\{\mu_k^N,
\Psi_k^N(\y)\}$ of the following exterior Steklov problem:

\begin{subequations}  \label{eq:PsiN_def}
\begin{align}  \label{eq:PsiN_Laplace}
\Delta \Psi_k^N & = 0 \quad (\y \in \R^3_+), \\
\partial_n \Psi_k^N & = \mu_k^N \Psi_k^N  \quad (\y \in \Gamma), \\
\partial_n \Psi_k^N & = 0 \quad (\y \in \partial \R^3_+ \backslash \overline{\Gamma}), \\  \label{eq:PsiN_inf}
|\y|^2 |\nabla \Psi_k^N| & \to 0 \quad (|\y|\to\infty).
\end{align}
\end{subequations}
While the decay condition (\ref{eq:Psi_inf}) could be interpreted as a
Dirichlet condition at infinity, the decay of the gradient of an
eigenfunction in Eq. (\ref{eq:PsiN_inf}) can be understood as a
Neumann condition at infinity (that is reflected by the superscript
'N').  The latter condition was studied in
\cite{Henrici70,Fox83,Davis70,Troesch72,Miles72} in
the context of sloshing or ice-fishing problems in hydrodynamics (see
also \cite{Kozlov04,Levitin22} and references therein); moreover, the
eigenpairs $\{\mu_k^N, \Psi_k^N\}$ naturally appear in the asymptotic
analysis of the mixed Steklov-Neumann problem
\cite{Grebenkov25b,GrebWard25a}.  In Appendix \ref{sec:FEM}, we
present an efficient numerical tool to solve the spectral problem
(\ref{eq:PsiN_def}).

The distinction between the spectral problems (\ref{eq:Psi_def}) and
(\ref{eq:PsiN_def}) becomes clearer from the form of their asymptotic
decay at infinity.  For instance, in the exterior of a ball, a general
isotropic solution of the Laplace equation is a linear combination
$c_N + c_D/|\y|$ with arbitrary constants $c_N$ and $c_D$.  The
Dirichlet condition (\ref{eq:Psi_inf}) imposes $c_N = 0$, whereas the
Neumann condition (\ref{eq:PsiN_inf}) imposes $c_D = 0$.  For general
domains (including the case of flat patches), the chosen Dirichlet or
Neumann condition fixes the far-field behavior of the solution.

In analogy to the spectral problem (\ref{eq:Psi_def}), the spectrum of
the mixed Steklov-Neumann problem (\ref{eq:PsiN_def}) is discrete,
with a countable set of eigenvalues, which can be enumerated by the
index $k$ in increasing order
\begin{equation}
0 = \mu_0^N < \mu_1^N \leq \mu_2^N \leq \ldots \nearrow +\infty .
\end{equation}
The restrictions of eigenfunctions onto the patch,
$\Psi_k^N|_{\Gamma}$, form a complete orthonormal basis of
$L^2(\Gamma)$.  Note that the principal eigenvalue $\mu_0^N = 0$ is
associated to the constant eigenfunction $\Psi_0^N =
1/\sqrt{|\Gamma|}$, and all other eigenfunctions are orthogonal to it.

The following spectral expansion was derived in \cite{GrebWard25a}:
\begin{equation}  \label{eq:Cmu_bis}
\frac{1}{C(\mu)} = \frac{1}{C(\infty)} + \frac{2\pi}{\mu |\Gamma|} 
+ 2\pi \sum\limits_{k=1}^\infty \frac{[\Psi_k^N(\infty)]^2}{\mu_k^N + \mu} \,.
\end{equation}
In this subsection, we discuss the complementary insights onto the
reactive capacitance that can be gained from this expansion.

\subsection{Sigmoidal approximation and several bounds}
\label{sec:approx}

Neglecting the positive sum in Eq. (\ref{eq:Cmu_bis}), one gets
immediately the lower bound:
\begin{equation}  \label{eq:Cmu_approx0}
\frac{1}{C(\mu)} \geq \frac{1}{C(\infty)} + \frac{2\pi}{\mu |\Gamma|} = \frac{1}{C^{\rm app}(\mu)} \,.
\end{equation}
Rewriting the right-hand side of this bound as
\begin{equation}  \label{eq:Cmu_approx}
C^{\rm app}(\mu) = \frac{\mu \, C(\infty)}{\mu + 2\pi C(\infty)/|\Gamma|} \,,
\end{equation}
we retrieve the sigmoidal approximation for the reactive capacitance.
This relation was proposed in \cite{GrebWard25a} as an empirical
approximation that correctly reproduces both small-$\mu$ and
large-$\mu$ limits of $C(\mu)$.  In turn, it follows here directly
from the spectral expansion (\ref{eq:Cmu_bis}) by neglecting its last
term, thus allowing one to control its accuracy (see below).
Note that a similar approximation appeared in \cite{Chaigneau22} in
the analysis of the principal eigenvalue of the Laplace operator with
Robin boundary condition, as well as in \cite{Plunkett24,Cengiz24}.
In the special case of a circular patch, the approximation
(\ref{eq:Cmu_approx}) was shown to be accurate over the entire range
of $\mu$ within the maximal relative error of $4\%$
\cite{GrebWard25a}.  One of the aims of the present work consists in
verifying the accuracy of the sigmoidal approximation for patches of
arbitrary shape.  For this purpose, we need to compute numerically the
Steklov eigenvalues $\mu_k$ and the spectral weights $F_k$ (see
Sec. \ref{sec:numerics}).

To obtain an upper bound on $1/C(\mu)$, one can rewrite
Eq. (\ref{eq:Cmu_bis}) as
\begin{align}  \label{eq:Cmu_auxil1}
\frac{1}{C(\mu)} & = \frac{1}{C(\infty)} + \frac{2\pi}{\mu|\Gamma|} + 2\pi \sum\limits_{k=1}^\infty \frac{[\Psi_k^N(\infty)]^2}{\mu_k^N} \\
\nonumber 
& - 2\pi \mu \sum\limits_{k=1}^\infty \frac{[\Psi_k^N(\infty)]^2}{\mu_k^N(\mu_k^N + \mu)} \,.
\end{align}
Taking the limit $\mu\to 0$ and using the Taylor expansion
(\ref{eq:Cmu_Taylor}) with two terms, $C(\mu) \approx c_1 \mu - c_2
\mu^2 + O(\mu^3)$, one computes the third term in the right-hand
side of Eq. (\ref{eq:Cmu_auxil1}) as
\begin{align}  \nonumber
\frac{1}{C(\infty)} + 2\pi \sum\limits_{k=1}^\infty \frac{[\Psi_k^N(\infty)]^2}{\mu_k^N}
& = \lim\limits_{\mu\to 0} \biggl(\frac{1}{C(\mu)} - \frac{2\pi}{\mu|\Gamma|} \biggr) \\  \label{eq:sum_bound}
& = \frac{c_2}{c_1^2} = 2\pi \A_\Gamma ,
\end{align}
where we used $c_1 = |\Gamma|/(2\pi)$ and expressed $c_2$ in terms of
the constant $\A_\Gamma$ defined in Eq. (\ref{eq:A_def}).  As a
consequence, Eq. (\ref{eq:Cmu_auxil1}) reads
\begin{equation}   \label{eq:Cmu_bis2}
\frac{1}{C(\mu)} = 2\pi \A_\Gamma + \frac{2\pi}{\mu|\Gamma|} 
- 2\pi \mu \sum\limits_{k=1}^\infty \frac{[\Psi_k^N(\infty)]^2}{\mu_k^N(\mu_k^N + \mu)} \,,
\end{equation}
that yields the upper bound
\begin{equation}   \label{eq:Cmu_lower2}
\frac{1}{C(\mu)} \leq 2\pi \A_\Gamma + \frac{2\pi}{\mu|\Gamma|}  \,.
\end{equation}
The two bounds can be written together as
\begin{equation}   \label{eq:Cmu_bounds}
\frac{1}{C(\infty)} \leq \frac{1}{C(\mu)} - \frac{2\pi}{\mu|\Gamma|} \leq 2\pi \A_\Gamma .
\end{equation}
We also note that the sum evaluated in Eq. (\ref{eq:sum_bound}) gives
the upper bound on the absolute error of the sigmoidal approximation
(\ref{eq:Cmu_approx}):
\begin{align}  
\frac{1}{C(\mu)} - \frac{1}{C^{\rm app}(\mu)} & =  2\pi \sum\limits_{k=1}^\infty \frac{[\Psi_k^N(\infty)]^2}{\mu_k^N + \mu}  \\  \nonumber
& \leq 2\pi \sum\limits_{k=1}^\infty \frac{[\Psi_k^N(\infty)]^2}{\mu_k^N} = 2\pi \A_\Gamma - \frac{1}{C(\infty)} \,.
\end{align}
Dividing this relation by $C^{\rm app}(\mu)$, we get the maximal bound
on the relative error:
\begin{equation}     \label{eq:Capprox_error}
\frac{C^{\rm app}(\mu) - C(\mu)}{C(\mu)}  \leq  \frac{\Emax}{1 + 2\pi C(\infty)/(\mu |\Gamma|)} \,,
\end{equation}
with
\begin{equation}   \label{eq:Emax}
\Emax = 2\pi \A_\Gamma C(\infty) - 1.
\end{equation}
One sees that the upper bound monotonously increases with $\mu$ and
reaches the maximum $\Emax$ at $\mu = \infty$.  However, the actual
relative error decreases in the limit $\mu\to \infty$ because $C(\mu)$
approaches to its limit $C(\infty)$, and the sigmoidal approximation
correctly captures this behavior.  As a consequence, the value $\Emax$
is expected to be a conservative over-estimation.  Using the spectral
expansions (\ref{eq:Cmu_Taylor}, \ref{eq:Fkmuk_sum}), one can
represent the maximal bound as
\begin{equation}
\Emax = \biggl(\sum\limits_{k=0}^\infty \mu_k F_k\biggr) \biggl(\sum\limits_{k=0}^\infty \frac{F_k}{\mu_k}\biggr) - 1 .
\end{equation}
According to Eq. (\ref{eq:Fk_sum}), one sees that if $F_0$ is close to
$1$, the dominant contributions to two sums are $\mu_0 F_0$ and
$F_0/\mu_0$ respectively, and the maximal bound is close to $0$.

The bounds (\ref{eq:Cmu_bounds}) can be further improved.  For
instance, rewriting Eq. (\ref{eq:Cmu_bis2}) as
\begin{align}   \nonumber
\frac{1}{C(\mu)} & = 2\pi \A_\Gamma + \frac{2\pi}{\mu|\Gamma|} 
- 2\pi \mu \sum\limits_{k=1}^\infty \frac{[\Psi_k^N(\infty)]^2}{[\mu_k^N]^2} \\   \label{eq:Cmu_bis3}
& + 2\pi \mu^2 \sum\limits_{k=1}^\infty \frac{[\Psi_k^N(\infty)]^2}{[\mu_k^N]^2(\mu_k^N + \mu)} \,,
\end{align}
one can neglect the last sum to obtain the improved lower bound on
$1/C(\mu)$:
\begin{equation}
\frac{1}{C(\mu)} \geq 2\pi \A_\Gamma + \frac{2\pi}{\mu|\Gamma|} - \mu B, 
\end{equation}
where the constant
\begin{equation}
B = 2\pi \sum\limits_{k=1}^\infty \frac{[\Psi_k^N(\infty)]^2}{[\mu_k^N]^2} = \frac{c_2^2 - c_1 c_3}{c_1^3}
\end{equation}
was expressed in terms of $c_i$ from Eq. (\ref{eq:cn_int}) via the
Taylor expansion (\ref{eq:Cmu_Taylor}).  Repeating this trick, one can
get more and more accurate bounds that include, however, higher-order
Taylor coefficients $c_n$.

\subsection{Probabilistic interpretations} 
\label{sec:prob}

From a physical point of view, the reactive capacitance is
proportional to the flux of particles diffusing from infinity, as
stated earlier.  In this subsection, we provide several probabilistic
interpretations of $C(\mu)$ that will allow us to establish its
additional properties.

Let $\W_t$ be the standard Brownian motion in $\R^3$ with diffusion
coefficient $D$, that starts from a fixed point $\y$ at time $0$.  The
reflected Brownian motion $\X_t$ in the upper half-space $\R^3_+$ can
simply be obtained by reflecting any random path $\W_t$ with respect
to the horizontal plane $y_3 = 0$; in other words, $\X_t = (W_t^1,
W_t^2, |W_t^3|)$, where $W_t^i$ are three independent components of
$\W_t$.  To incorporate partial reactivity of the patch $\Gamma$, we
follow the encounter-based approach \cite{Grebenkov20}.  For this
purpose, we introduce the boundary local time $\ell_t$ on $\Gamma$ as
the limit of the rescaled residence time $\ell_t^{(\sigma)}$ of
$\X_t$, up to time $t$, in a thin boundary layer of thickness $\sigma$
near $\Gamma$:
\begin{equation}
\ell_t = \lim\limits_{\sigma\to 0} \ell_t^{(\sigma)},   \quad
\ell_t^{(\sigma)} = \frac{D}{\sigma} \int\limits_0^t \Theta(\sigma - |\X_{t'} - \Gamma|) dt' ,
\end{equation}
where $\Theta(z)$ is the Heaviside step function: $\Theta(z) = 1$ for
$z > 0$, and $0$ otherwise.  We note that both $\ell_t$ and
$\ell_t^{(\sigma)}$ have units of length, despite their name.  The
boundary local time $\ell_t$ can also be interpreted as the rescaled
number of arrivals of $\X_t$ onto $\Gamma$ up to time $t$.  By
construction, $\ell_t$ is a nondecreasing stochastic process, with
$\ell_0 = 0$.  Since the reflected Brownian motion in $\R^3_+$ is
transient, its ability to revisit the patch is limited so that the
boundary local time $\ell_t$ has a finite (random) limit $\ell_\infty$
as $t\to \infty$.  Let us denote its probability density by
$\rho_\infty(\ell|\y)$.

At each arrival onto the patch $\Gamma$, the diffusing particle
modeled by reflected Brownian motion may react and thus disappear.
For a constant reactivity parameter $\mu$, the reaction event is
triggered when the number of arrivals (i.e., failed reaction
attempts), represented by $\ell_t$, exceeds a random threshold
$\hat{\ell}$ that obeys the exponential distribution: $\P\{\hat{\ell}
> \ell\} = e^{-\mu \ell}$ (see \cite{Grebenkov20} for details).  In
this way, the first-reaction time (FRT) on the patch $\Gamma$ is
defined as the first-crossing time of the threshold $\hat{\ell}$ by
$\ell_t$: $\tau = \inf\{ t > 0 ~:~ \ell_t > \hat{\ell}\}$.  Since the
reflected Brownian motion in $\R^3_+$ is transient, there is a finite
probability of escaping to infinity without reacting on the patch; in
this case, the FRT is infinite: $\tau = +\infty$.  In turn, the
probability of the reaction event, $\P_{\y}\{ \tau < +\infty\}$, is
known to satisfy Eq. (\ref{eq:wi_def}) and is thus given by
$w(\y;\mu)$.  We conclude that the auxiliary function $w(\y;\mu)$
admits a simple interpretation as the probability of reaction on the
patch $\Gamma$ with the reactivity parameter $\mu$, when the starting
point is $\y$.  In turn, $1-w(\y;\mu)$ is the escape probability
without reaction.  We also note that the reaction time $\tau$ is
finite if and only if the limit $\ell_\infty$ is above the threshold
$\hat{\ell}$, i.e., the reaction event $\{ \tau < + \infty\}$ is
identical to $\{ \ell_\infty > \hat{\ell}\}$.  As a consequence, we
get
\begin{align} \label{eq:w_moments}
w(\y;\mu) & = \P_{\y}\{ \tau < +\infty\} = \P_{\y} \{ \ell_\infty > \hat{\ell}\}  \\  \nonumber
& = \int\limits_0^\infty \underbrace{\mu e^{-\mu \ell}}_{\textrm{PDF of}~\hat{\ell}} ~ \cdot ~ 
\underbrace{\P_{\y} \{ \ell_\infty > \ell \}}_{\int\nolimits_{\ell}^\infty \rho_\infty(\ell'|\y) d\ell'} d\ell  \\  \nonumber
& = 1 - \int\limits_0^\infty  e^{-\mu \ell} \, \rho_\infty(\ell|\y) \, d\ell = 1 - \E_{\y}\{ e^{-\mu \ell_\infty}\},
\end{align}
where we integrated by parts.  We conclude that $1 - w(\y;\mu)$ is the
moment-generating function of the random variable $\ell_\infty$.

In the last step, we consider the situation when the starting position
of the particle is not fixed but randomly chosen on the patch with the
uniform density.  The associated escape probability reads
\begin{equation}   \label{eq:Puniform}
P_{u}(\mu) =  \int\limits_{\Gamma} (1-w(\y;\mu)) \frac{d\y}{|\Gamma|}
 = \frac{2\pi C(\mu)}{\mu |\Gamma|} \,,
\end{equation}
where we used the boundary condition (\ref{eq:wi_Robin}), the
divergence theorem, and the asymptotic behavior (\ref{eq:wi_inf}) at
infinity (the subscript $u$ refers to the uniform distribution of
the starting point on $\Gamma$).  Inverting this relation, one can
express the reactive capacitance $C(\mu)$ in terms of the escape
probability on the patch.  In turn, using Eq. (\ref{eq:w_moments}), we
can relate $C(\mu)$ to the moment-generating function:
\begin{equation}  \label{eq:Ecirc_mu}
\E_{u}\{ e^{-\mu \ell_\infty}\} = \frac{2\pi C(\mu)}{\mu |\Gamma|} \,,
\end{equation}
where 
\begin{equation}
\E_{u}\{ e^{-\mu \ell_\infty}\} = \int\limits_{\Gamma} \E_{\y}\{ e^{-\mu \ell_\infty}\} \, \frac{d\y}{|\Gamma|} 
\end{equation}
corresponds to the uniform starting point on the patch.  Comparing
this expression with the Taylor expansion (\ref{eq:Cmu_Taylor}), we
get an interesting probabilistic interpretation of the Taylor
coefficients $c_n$ as the moments of the boundary local time
$\ell_\infty$:
\begin{equation}
\E_{u}\{ \ell_\infty^n \} = \frac{2\pi}{|\Gamma|} c_{n+1}  \quad (n = 0,1,\ldots). 
\end{equation}
For instance, we retrieve $c_1 = |\Gamma|/(2\pi)$ at $n = 0$, whereas
$c_2$ is proportional to the mean boundary local time.   As a
consequence, we get
\begin{equation}
\A_\Gamma = \frac{\E_{u}\{ \ell_\infty \}}{|\Gamma|} \,.
\end{equation}
Curiously, the same constant $\A_\Gamma$ was shown to determine the
asymptotic behavior of the variance of the boundary local time
$\ell_t$ on a small patch for restricted diffusion in a bounded domain
\cite{Grebenkov25b}.
Moreover, substituting the spectral expansion (\ref{eq:Cmu}) into
Eq. (\ref{eq:Ecirc_mu}) and performing the inverse Laplace transform
with respect to $\mu$, one gets the probability density of
$\ell_\infty$:
\begin{equation}
\rho_\infty(\ell|u) = \sum\limits_{k=0}^\infty \mu_k F_k e^{-\mu_k \ell} 
\end{equation}
(see \cite{Grebenkov20,Grebenkov23a} for further discussions on such
spectral representations).

An alternative probabilistic interpretation of $C(\mu)$ can be deduced
by multiplying Eq. (\ref{eq:wi_eq}) by $w(\y;\infty)$,
Eq. (\ref{eq:wi_eq}) with $\mu = \infty$ by $w(\y;\mu)$, subtracting
and integrating these equations over $\y\in\R^3_+$, and using the
Green's formula, boundary conditions and the behavior at infinity:
\begin{equation}
\int\limits_{\Gamma} w(\y;\mu) \partial_n w(\y;\infty) d\y = \int\limits_{\Gamma} \partial_n w(\y;\mu) d\y = 2\pi C(\mu).
\end{equation}
Dividing this expression by $2\pi C(\infty)$, we get
\begin{equation}  \label{eq:Preact1}
P_h(\mu) = \int\limits_{\Gamma} w(\y;\mu) h(\y) d\y = \frac{C(\mu)}{C(\infty)} \,,
\end{equation}
where 
\begin{equation} \label{eq:h_def}
h(\y) = \frac{1}{2\pi C(\infty)} \partial_n w(\y;\infty)|_{\Gamma} 
\end{equation}
can be interpreted as the (conditional) harmonic measure density for a
particle arriving from infinity, i.e., $h(\y) d\y$ is the probability
of the first arrival onto the patch $\Gamma$ in a $d\y$ vicinity of
the point $\y\in \Gamma$.  In other words, $h(\y)$ is the normalized
flux density of particles from infinity.  In this light, the integral
in Eq. (\ref{eq:Preact1}) can be interpreted as the probability of
reaction for the process that has managed to arrive from infinity onto
the patch.

\subsection{Domain monotonicity}

The electrostatic capacitance $C(\infty)$ is known to satisfy the
domain monotonicity property (see \cite{Landkof,Adams} for details):
\begin{equation}  \label{eq:monotonicity_inf}
\textrm{If} ~ \Gamma_1 \subset \Gamma_2  \quad \Rightarrow \quad C_1(\infty) \leq C_2(\infty),
\end{equation}
i.e., the larger patch has the larger trapping ability.  This property
is an immediate consequence of the classical variational formulation
of the electrostatic capacitance, which can be defined for compact
sets (including the case of flat patches) as 
% see Adams and Hedberg, p. 18
%
\begin{equation}
C(\infty) = \frac{1}{4\pi} \inf\biggl\{ \int\limits_{\R^3} |\nabla u|^2 d\y ~:~ u\in C_0^\infty(\R^3), ~  u|_\Gamma \geq 1 \biggr\},
\end{equation} 
where $C^\infty_0(\R^3)$ is the space of smooth functions in $\R^3$
with a compact support, and the factor $4\pi$ is included to ensure
our convention that the electrostatic capacitance of a ball of radius
$R$ is $R$; a similar definition can be written for the upper
half-space.  Note that the functional space $C_0^\infty$ can be
replaced by the Sobolev space $H^1(\R^3)$ with appropriate
restrictions.

It is intuitively expected that the same statement holds for partially
reactive patches, i.e., a larger patch captures diffusing particles
more efficiently.  In this subsection, we establish this domain
monotonicity property and some other monotonicity relations.

\subsubsection*{First monotonicity relation}

For this purpose, we first derive another representation of $C(\mu)$
by setting $\bar{w}(\y;\mu) = 1 - w(\y;\mu)$, which satisfies 
\begin{subequations}  \label{eq:wibar_def}
\begin{align}  \label{eq:wibar_eq}
\Delta \bar{w} & = 0 \quad (\y \in \R^3_+), \\  \label{eq:wibar_Robin}
\partial_n \bar{w} + \mu \bar{w} & = 0  \quad (\y \in \Gamma), \\  
\partial_n \bar{w} & = 0 \quad (\y \in \partial \R^3_+ \backslash \overline{\Gamma}), \\  \label{eq:wibar_inf}
\bar{w} & \sim 1 - \frac{C(\mu)}{|\y|} + \O(|\y|^{-2}) \quad (|\y|\to\infty).
\end{align}
\end{subequations}
Multiplying this equation by $\bar{w}$ and integrating over $\R^3_+$,
we get
\begin{align*}
0 & = \int\limits_{\R^3_+} \bar{w} \,\Delta \bar{w} \, d\y = - \int\limits_{\R^3_+} |\nabla \bar{w}|^2\, d\y 
+ \int\limits_{\Gamma} \bar{w} \,\partial_n \bar{w} \, d\y  \\
& = - \int\limits_{\R^3_+} |\nabla \bar{w}|^2\, d\y - \mu \int\limits_{\Gamma} \bar{w}^2\, d\y  + 2\pi C(\mu),
\end{align*}
where we used the Green's formula to integrate by parts, the Robin
boundary condition (\ref{eq:wibar_Robin}), and the behavior
(\ref{eq:wibar_inf}) at infinity.  As a consequence, we have
\begin{equation}
C(\mu) = \frac{1}{2\pi} \biggl\{ \int\limits_{\R^3_+} |\nabla w|^2 \, d\y + \mu \int\limits_{\Gamma} (1- w)^2 \,d\y \biggr\} ,
\end{equation}
where we returned to the original function $w(\y;\mu)$.

Following the standard arguments, one can recast this relation in a
variational form:
\begin{equation}  \label{eq:Cmu_var1}
C(\mu) = \frac{1}{2\pi} \inf\limits_{u\in C^\infty_0(\R^3_+)} 
\biggl\{ \int\limits_{\R^3_+} |\nabla u|^2 d\y + \mu \int\limits_{\Gamma} (1-u)^2\, d\y \biggr\} .
\end{equation}
The variational formulation immediately implies the domain
monotonicity property for the reactive capacitance:
\begin{equation}  \label{eq:monotonicity}
\textrm{If} ~ \Gamma_1 \subset \Gamma_2  \quad \Rightarrow \quad C_1(\mu) \leq C_2(\mu)~~~ \forall~ \mu \geq 0,
\end{equation}
where $C_1(\mu)$ and $C_2(\mu)$ refer to the reactive capacitances of
$\Gamma_1$ and $\Gamma_2$, respectively.  Indeed, as one searches the
infimum over the same functional space, any reduction of the patch
diminishes the second term in Eq. (\ref{eq:Cmu_var1}) and thus the
associated $C_i(\mu)$.  This is an extension of
Eq. (\ref{eq:monotonicity_inf}) to a finite reactivity $\mu > 0$.

\subsubsection*{Second monotonicity relation}

In the same vein, one can multiply Eq. (\ref{eq:wi_eq}) by $w(\y;\mu)$
and then integrating over $\R^3_+$ to get:
\begin{align*}
0 & = \int\limits_{\R^3_+} w \,\Delta w \, d\y = - \int\limits_{\R^3_+} |\nabla w|^2\, d\y + \int\limits_{\Gamma} w \,\partial_n w \, d\y  \\
& = - \int\limits_{\R^3_+} |\nabla w|^2\, d\y + \mu (\mu|\Gamma| - 2\pi C(\mu)) - \mu \int\limits_{\Gamma} w^2\, d\y,
\end{align*}
where we used again the Green's formula, the boundary condition
(\ref{eq:wi_Robin}), and the behavior (\ref{eq:wi_inf}) at infinity
such that
\begin{equation}  \label{eq:Cmu_int}
\mu \int\limits_{\Gamma} (1 - w) d\y = \int\limits_{\Gamma} \partial_n w\, d\y = 2\pi C(\mu).
\end{equation}
We have thus
\begin{equation}
\delta C(\mu) = \frac{1}{2\pi \mu} \biggl\{ \int\limits_{\R^3_+} |\nabla w|^2\, d\y 
+ \mu \int\limits_{\Gamma} w^2\, d\y\biggr\},
\end{equation}
with a shortcut notation
\begin{equation}
\delta C(\mu) = \frac{\mu |\Gamma|}{2\pi} - C(\mu).
\end{equation}
The variational form reads
\begin{equation}  \label{eq:Cmu_var2}
\delta C(\mu) = \frac{1}{2\pi \mu} \inf\limits_{u\in C^\infty_0(\R^3_+)} 
\biggl\{ \int\limits_{\R^3_+} |\nabla u|^2 d\y + \mu \int\limits_{\Gamma} u^2\, d\y \biggr\} 
\end{equation}
and thus yields the second domain monotonicity property:
\begin{equation}  \label{eq:monotonicity2}
\textrm{If} ~ \Gamma_1 \subset \Gamma_2  \quad \Rightarrow \quad \delta C_1(\mu) \leq \delta C_2(\mu)~~~ \forall~ \mu \geq 0.
\end{equation}

This property could alternatively be obtained from the probabilistic
interpretation.  For any fixed $\y$ and $\mu$, the boundary local time
on the larger patch $\Gamma_2$ is greater than or equal to the
boundary local time on the smaller patch, so that the first-reaction
time on $\Gamma_2$ is smaller and thus the particle has more chances
to react with $\Gamma_2$ before escaping to infinity.  As a
consequence, one has
\begin{equation}
\textrm{If} ~ \Gamma_1 \subset \Gamma_2  \quad \Rightarrow \quad  0\leq w_1(\y;\mu) \leq w_2(\y;\mu)  ~~~ 
\left(\mu \geq 0, \atop \y\in\R^3_+ \right).
\end{equation}  
The integral of this inequality over $\Gamma_1$ yields then
\begin{equation*}
\int\limits_{\Gamma_1} w_1(\y;\mu) d\y \leq \int\limits_{\Gamma_1} w_2(\y;\mu) d\y \leq \int\limits_{\Gamma_2} w_2(\y;\mu) d\y ,
\end{equation*}
from which Eq. (\ref{eq:Cmu_int}) implies
\begin{equation}
C_2(\mu) \leq C_1(\mu) + \frac{|\Gamma_2| - |\Gamma_1|}{2\pi} \mu \,.
\end{equation}

\subsubsection*{Third monotonicity relation}

Next, we establish another monotonicity relation.  For this purpose,
let us consider the auxiliary function
\begin{equation}
\tilde{w}(\y;\mu) = \frac{w(\y;\mu)}{C(\mu)} - \frac{w(\y;\infty)}{C(\infty)} \,,
\end{equation}
which satisfies, by construction,
\begin{subequations}  \label{eq:tildewi_def}
\begin{align}  \label{eq:tildewi_eq}
\Delta \tilde{w} & = 0 \quad (\y \in \R^3_+), \\  \label{eq:tildewi_Robin}
\partial_n \tilde{w} + \mu \tilde{w} & 
= \frac{\mu}{C(\mu)} - \frac{\mu}{C(\infty)} - 2\pi h(\y)  \quad (\y \in \Gamma), \\
\partial_n \tilde{w} & = 0 \quad (\y \in \partial \R^3_+ \backslash \overline{\Gamma}), \\ 
\tilde{w} & \sim \O(|\y|^{-2}) \quad (|\y|\to\infty),
\end{align}
\end{subequations}
where $h(\y)$ was defined in Eq. (\ref{eq:h_def}).  Since $\tilde{w}$
decays faster than $1/|\y|$, the divergence theorem implies
\begin{equation}  \label{eq:_int_dtildewi}
\int\limits_{\Gamma} \partial_n \tilde{w} \, d\y = 0 ,
\end{equation}
from which the boundary condition (\ref{eq:tildewi_Robin}) yields
\begin{equation}  \label{eq:_int_tildewi}
\int\limits_{\Gamma} \tilde{w}(\y;\mu)d\y = \frac{|\Gamma|}{C(\mu)} - \frac{|\Gamma|}{C(\infty)} - \frac{2\pi}{\mu} \,,
\end{equation}
where we used that $\int\nolimits_{\Gamma} h(\y)\, d\y = 1$.
Multiplying Eq. (\ref{eq:tildewi_eq}) by $\tilde{w}$ and integrating
over $\R^3_+$, we get
\begin{equation*} 
0 = \int\limits_{\R^3_+} \tilde{w} \, \Delta \tilde{w} \, d\y = - \int\limits_{\R^3_+} |\nabla \tilde{w}|^2 d\y
+ \int\limits_{\Gamma} \tilde{w} \, \partial_n \tilde{w} \, d\y.
\end{equation*} 
Substituting $\partial_n \tilde{w}$ from the boundary condition
(\ref{eq:tildewi_Robin}), we have
\begin{align*}
0 & = - \int\limits_{\R^3_+} |\nabla \tilde{w}|^2 \, d\y
- \mu \int\limits_{\Gamma_i} \tilde{w}^2 \, d\y  \\
& + \biggl(\frac{\mu}{C(\mu)} - \frac{\mu}{C(\infty)}\biggr) \int\limits_{\Gamma} \tilde{w} \, d\y 
 -  2\pi \int\limits_{\Gamma} \tilde{w}\, h  \, d\y\, .
\end{align*} 
To evaluate the last term, we multiply Eq. (\ref{eq:tildewi_eq}) by
$w(\y;\infty)$, multiply Eq. (\ref{eq:wi_eq}) with $\mu = \infty$
by $\tilde{w}(\y;\mu)$, subtract them, integrate over $\R^3_+$, use
the Green's formula, boundary conditions and the decay at infinity to
get
\begin{equation}
\int\limits_{\Gamma} \tilde{w}\, h \, d\y  = \int\limits_{\Gamma} \underbrace{w(\y;\infty)}_{=1} \partial_n \tilde{w} \, d\y = 0 ,
\end{equation}
where we used Eq. (\ref{eq:_int_dtildewi}).  Finally, substituting
Eq. (\ref{eq:_int_tildewi}), we obtain
\begin{align} \nonumber
& \mu |\Gamma| \biggl(\frac{1}{C(\mu)} - \frac{1}{C(\infty)}\biggr) \biggl(\frac{1}{C(\mu)} - \frac{1}{C(\infty)} 
- \frac{2\pi}{\mu |\Gamma|}\biggr)  \\   \label{eq:Cdiff_integral}
& = \int\limits_{\R^3_+} |\nabla \tilde{w}|^2\, d\y + \mu \int\limits_{\Gamma} \tilde{w}^2 \, d\y.
\end{align}
Using again the variational form, we obtain the following inequality
\begin{align} \nonumber
& |\Gamma_1| \biggl(\frac{1}{C_1(\mu)} - \frac{1}{C_1(\infty)}\biggr) \biggl(\frac{1}{C_1(\mu)} - \frac{1}{C_1(\infty)} 
- \frac{2\pi}{\mu |\Gamma_1|}\biggr)   \\     \label{eq:monotonicity3}
& \leq |\Gamma_2| \biggl(\frac{1}{C_2(\mu)} - \frac{1}{C_2(\infty)}\biggr) \biggl(\frac{1}{C_2(\mu)} - \frac{1}{C_2(\infty)} 
- \frac{2\pi}{\mu |\Gamma_2|}\biggr)   
\end{align}
if $\Gamma_1 \subset \Gamma_2$.

\subsubsection*{Fourth monotonicity relation}

From the last inequality, let us deduce a weaker yet more appealing
monotonicity property.  Denoting $\eta_i = |\Gamma_i| (1/C_i(\mu) -
1/C_i(\infty))$, we rewrite Eq. (\ref{eq:monotonicity3}) as
\begin{equation}
\frac{\eta_1 (\eta_1 - 2\pi/\mu)}{|\Gamma_1|}  \leq  \frac{\eta_2 (\eta_2 - 2\pi/\mu)}{|\Gamma_2|} \,,
\end{equation}
or, equivalently,
\begin{equation}
\frac{(\eta_1 - \pi/\mu)^2 - \pi^2/\mu^2}{|\Gamma_1|}  \leq  \frac{(\eta_2 - \pi/\mu) - \pi^2/\mu^2}{|\Gamma_2|} \,.
\end{equation}
Since $|\Gamma_1| \leq |\Gamma_2|$, multiplication of two inequalities
yields, after few simplifications,
\begin{equation}
|\eta_1 - \pi/\mu|  \leq  |\eta_2 - \pi/\mu| \,.
\end{equation}
Since $\eta_i \geq 2\pi/\mu$ due to Eq. (\ref{eq:Cmu_bis}), we
conclude that $\eta_1 \leq \eta_2$, i.e.,
\begin{equation}  \label{eq:monotonicity4}
|\Gamma_1| \biggl(\frac{1}{C_1(\mu)} - \frac{1}{C_1(\infty)}\biggr)\leq
|\Gamma_2| \biggl(\frac{1}{C_2(\mu)} - \frac{1}{C_2(\infty)}\biggr).
\end{equation}
This inequality allows one to control the large-reactivity behavior of
the reactive capacitance.

\subsection{Large-reactivity limit}

When the reactivity parameter $\mu$ goes to infinity, the reactive
capacitance $C(\mu)$ approaches its limit $C(\infty)$.  However, the
behavior of the reactive capacitance is not analytic at infinity.  In
fact, according to Eq. (\ref{eq:Ci_def}), the reactive capacitance is
the integral of the flux density, which may diverge at the boundary of
the patch in the Dirichlet setting.  This is well-known for an
elliptic patch with semiaxes $a \leq b$, for which the flux density is
given by (see, e.g., \cite{Strieder09})
\begin{equation}  \label{eq:q_elliptic}
\partial_n w(\y;\infty)|_{\Gamma} = \frac{\bigl(1 - y_1^2/a^2 - y_2^2/b^2\bigr)^{-1/2}}{a K(\sqrt{1-a^2/b^2})}  \,,
\end{equation}
where $K(z)$ is the complete elliptic integral of the first kind:
\begin{equation}  \label{eq:ellipticK}
K(z) = \int\limits_0^{\pi/2} \frac{d\theta}{\sqrt{1 - z^2 \sin^2 \theta}} \,.
\end{equation}
The effect of this singular behavior onto the mean first-reaction time
on a circular patch was analyzed in \cite{Guerin23} by using a
Wiener-Hopf integral equation.  Relying on this analysis, the
large-$\mu$ asymptotic behavior of the reactive capacitance for the
unit disk was deduced in \cite{GrebWard25a}:
\begin{equation}  \label{eq:C_mularge_disk}
C_\circ(\mu) \approx C_\circ(\infty) - \frac{2(\ln (2\mu) + \gamma + 1)}{\pi^2 \mu}  \quad (\mu \to \infty),
\end{equation}
where $\gamma \approx 0.5772\ldots$ is the Euler constant,
$C_\circ(\infty) = 2/\pi$, and subscript $\circ$ refers to the unit
disk.  The presence of the logarithm makes this expression
non-analytic.  Does this behavior hold for noncircular patches?

While a systematic analysis of this problem for arbitrary patches
remains an open problem, we argue here that the non-analytic
large-$\mu$ behavior is a generic feature of flat patches.  For this
purpose, we first rewrite Eq. (\ref{eq:C_mularge_disk}) as
\begin{equation}  \label{eq:C_mularge_disk2}
\frac{1}{C_\circ(\mu)} - \frac{1}{C_\circ(\infty)} \approx \frac{\ln(2\mu) + \gamma + 1}{2\mu}  \quad (\mu \to \infty).
\end{equation}
This relation can now be used as a bound according to the monotonicity
relation (\ref{eq:monotonicity4}).  In fact, for any given patch
$\Gamma$, one can inscribe a small disk $\Gamma_0$ of radius $r_0$.  As
$\Gamma_0 \subset \Gamma$, one has
\begin{equation}  
\frac{\pi r_0}{|\Gamma|} \biggl(\frac{1}{C_\circ(r_0 \mu)} - \frac{1}{C_\circ(\infty)}\biggr)\leq
\frac{1}{C(\mu)} - \frac{1}{C(\infty)} \,,
\end{equation}
where we used the rescaling relation (\ref{eq:Cmu_scaling}).  Even
though the asymptotic relation (\ref{eq:C_mularge_disk2}) is not a
bound, one can add an additional prefactor to turn it into a bound for
large $\mu$, i.e.,
\begin{equation}   \label{eq:Cmu_log}
\beta \frac{\ln(2r_0 \mu) + \gamma + 1}{2\mu} \leq \frac{1}{C(\mu)} - \frac{1}{C(\infty)} \quad (\mu \to \infty) ,
\end{equation}
with some constant $\beta > 0$.  We conclude that the difference in
the right-hand side cannot vanish faster than $\ln(\mu)/\mu$, i.e.,
the logarithmic singularity is also expected at infinity, as for the
circular patch.  A further work is needed to characterize this
behavior more accurately.
  
The non-analytic behavior of $C(\mu)$ is tightly related to the
divergence of the sum of coefficients $[V_k^N(\infty)]^2$.  In fact,
the above asymptotic behavior, together with the spectral expansion
(\ref{eq:Cmu_bis}), implies that
\begin{equation}  
\mu \biggl(\frac{1}{C(\mu)} - \frac{1}{C(\infty)}\biggr) 
= 2\pi \sum\limits_{k=0}^\infty \frac{\mu [\Psi_k^N(\infty)]^2}{\mu_k^N + \mu}  \propto \ln(\mu) \to \infty ,
\end{equation}
as $\mu \to \infty$.  This divergence suggests that their partial sum
also diverges logarithmically:
\begin{equation}
\sum\limits_{k=1}^K [\Psi_k^N(\infty)]^2 \propto \ln(K)  \qquad (K\to \infty).
\end{equation}
It is instructive to give an alternative insight.  For this purpose,
let us multiply Eq. (\ref{eq:PsiN_Laplace}) by $w(\y;\infty)$,
Eq. (\ref{eq:wi_eq}) by $\Psi_k^N(\y)$, subtract them, integrate over
$\R^3_+$ and use the boundary conditions and the asymptotic behavior
at infinity to get
\begin{equation}
\Psi_k^N(\infty) = \int\limits_{\Gamma} h(\y) \, \Psi_k^N(\y) d\y ,
\end{equation}
where $h(\y)$ is defined by Eq. (\ref{eq:h_def}).  Since
$\{\Psi_k^N\}$ form a complete basis of $L^2(\Gamma)$, one can
formally write
\begin{align*}
\sum\limits_{k=0}^\infty [\Psi_k^N(\infty)]^2 & = \int\limits_{\Gamma} d\y_1 \, h(\y_1) \int\limits_{\Gamma} d\y_2 \, h(\y_2) \\
& \times \underbrace{\sum\limits_{k=0}^\infty \Psi_k^N(\y_1) \Psi_k^N(\y_2)}_{=\delta(\y_1-\y_2)} 
 = \int\limits_{\Gamma} h^2(\y) \, d\y = \infty .
\end{align*} 
In other words, the squared harmonic measure density diverges
logarithmically due to the edge singularity.  This is exemplified by
Eq. (\ref{eq:q_elliptic}) that provides an explicit form of $h(\y)$
for elliptic patches.  In summary, even though the coefficients
$[\Psi_k^N(\infty)]^2$ determine the relative contributions of the
Steklov eigenfunctions $\Psi_k^N$ to the reactive capacitance via
Eq. (\ref{eq:Cmu_bis}), in analogy to the coefficients $F_k$ in
Eq. (\ref{eq:Cmu}), the latter are summed to $1$ that facilitates
their interpretation as relative weights of the Steklov eigenfunctions
$\Psi_k$ to the reactive capacitance $C(\mu)$.

\subsection{Bounds for the principal eigenvalue}
\label{sec:bounds}

While there were numerous studies on isoperimetric inequalities for
Steklov eigenvalues in the conventional setting of interior domains
(see \cite{Levitin,Girouard17,Colbois24} and references therein), much
less is known about the exterior Steklov problem
\cite{Xiong23,Bundrock25}.  Moreover, the assumptions on the
regularity of the boundary that were used in former works, do not
necessarily hold for flat patches.  For instance, a circular patch can
be seen as the limit of thinning oblate spheroids, whose curvature
diverges (the edge singularity).  To fulfill this gap, we briefly
mention some inequalities that follow from our spectral expansions,
without pretending for rigorous proofs.

Since $\mu_k$ and $F_k$ are positive, one can use the spectral
expansion (\ref{eq:Fkmuk_sum}) to recover the Payne's upper bound for
the principal eigenvalue $\mu_0$ \cite{Payne56}:
\begin{equation}  \label{eq:mu0_upper}
\mu_0 \leq \frac{2\pi C(\infty)}{|\Gamma|}  \,.
\end{equation}
In a similar way, one can substitute $\mu_0 < \mu_k$ into the spectral
representations (\ref{eq:Cmu_Taylor}) of the Taylor coefficients $c_n$
to get
\begin{equation}  \label{eq:mu0_upper_cn}
\mu_0 \leq \biggl(\frac{|\Gamma|}{2\pi c_{n+1}}\biggr)^{1/n}  \,, \qquad n = 1,2,\ldots.
\end{equation}
Moreover, keeping only the first term in the spectral expansion
(\ref{eq:Cmu_Taylor}), one gets the {\it lower} bound:
\begin{equation}  \label{eq:mu0_lower_cn}
\mu_0 \geq \biggl(\frac{F_0 \, |\Gamma|}{2\pi c_{n+1}}\biggr)^{1/n}  \,, \qquad n = 1,2,\ldots,
\end{equation}
which is less explicit as it includes the spectral coefficient $F_0$.
For instance, one has for $n = 1$
\begin{equation}  \label{eq:mu0_boundsA}
\frac{F_0}{\A_\Gamma |\Gamma|} \leq \mu_0 \leq \frac{1}{\A_\Gamma |\Gamma|} \,,
\end{equation}
with $\A_\Gamma$ being defined in Eq. (\ref{eq:A_def}).
Alternatively, the inequality (\ref{eq:mu0_lower_cn}) can be seen as
an upper bound on $F_0$:
\begin{equation}  \label{eq:F0_bound}
F_0 \leq \frac{2\pi c_{n+1}\mu_0^n}{|\Gamma|}   \,, \qquad n = 1,2,\ldots,
\end{equation}
or, equivalently,
\begin{equation}  \label{eq:F0_bound2}
\int\limits_{\Gamma} \Psi_0 \, d\y \leq \sqrt{2\pi c_{n+1}\mu_0^n}   \,, \qquad n = 1,2,\ldots
\end{equation}

To illustrate these bounds, let us consider the circular patch of unit
radius, for which the first three Taylor coefficients were obtained in
\cite{GrebWard25a}:
\begin{equation}
c_1 = \frac12,  \quad c_2 = \frac{4}{3\pi}, \quad c_3 = \frac{4}{\pi^2}\int\limits_0^1 x E^2(x) dx \approx 0.3651,
\end{equation}
where $E(x)$ is the complete elliptic integral of the second kind.
Substituting the values of $C(\infty) = 2/\pi$, $c_2$ and $c_3$ into
Eqs. (\ref{eq:mu0_upper}, \ref{eq:mu0_upper_cn}), we get three upper
bounds: $1.263$ (Payne's bound), $1.178$ ($n = 1$), and $1.170$ ($n =
2$), whereas the numerical value of the principal eigenvalue is
$1.159$.  One sees that the bound becomes tighter as $n$ increases, as
expected from the spectral form (\ref{eq:Fkmuk_sum}).  Using the
numerical value $F_0 \approx 0.978$, we get the lower bounds $1.152$
and $1.157$ for $n=1$ and $n=2$, and the last bound is very close to
the numerical value of $\mu_0$.  We recall that $c_2$ and $c_3$ for
arbitrary patches are expressed via Eq. (\ref{eq:cn_int}) in terms of
the function $\omega_\Gamma(\x)$ from Eq. (\ref{eq:w_def}).  In
Appendix \ref{sec:wGamma}, we compute $\A_\Gamma$ for rectangular and
elliptic patches; in the latter case, we also compare the upper bounds
(\ref{eq:mu0_upper}) and (\ref{eq:mu0_boundsA}).

\subsection{Relation to the patch shape}

To gain further insights onto the relation between the reactive
capacitance of the patch and its geometric shape, one needs to solve
the exterior Steklov problem (\ref{eq:Psi_def}).  In Appendix
\ref{sec:FEM}, we develop an efficient numerical approach, which
relies on the reformulation (\ref{eq:Psi_integral}) of the original
problem.  This is a finite-element method, which employs linear basis
functions on a triangular mesh on the patch.  The key step to achieve
good accuracy and rapidity of the method is a semi-analytical
representation of the integral kernel $\G(\y,\y')$ on these basis
functions.  The accuracy of the method, which is controlled by the
mesh size, was validated by considering the circular patch, for which
an alternative very accurate approach based on oblate spheroidal
coordinates is available \cite{Grebenkov24}.

Using this numerical tool, we aim at answering the following
questions:

(i) How does anisotropy of the patch affect the Steklov eigenvalues
and eigenfunctions?

(ii) Does the principal eigenfunction $\Psi_0$ always provide the
dominant contribution to the reactive capacitance?

(iii) What is the accuracy of the empirical approximation
(\ref{eq:Cmu_approx}) for different patches?

\section{Numerical results}
\label{sec:numerics}

\subsection{Steklov eigenpairs}

Since the Steklov eigenfunctions determine the reactive capacitance,
it is instructive to first inspect their behavior for different
patches.  We start with the circular patch as a standard benchmark
example, and then explore other patch shapes.

\subsubsection*{Circular patch}

\begin{figure}
\begin{center}
\includegraphics[width=\columnwidth]{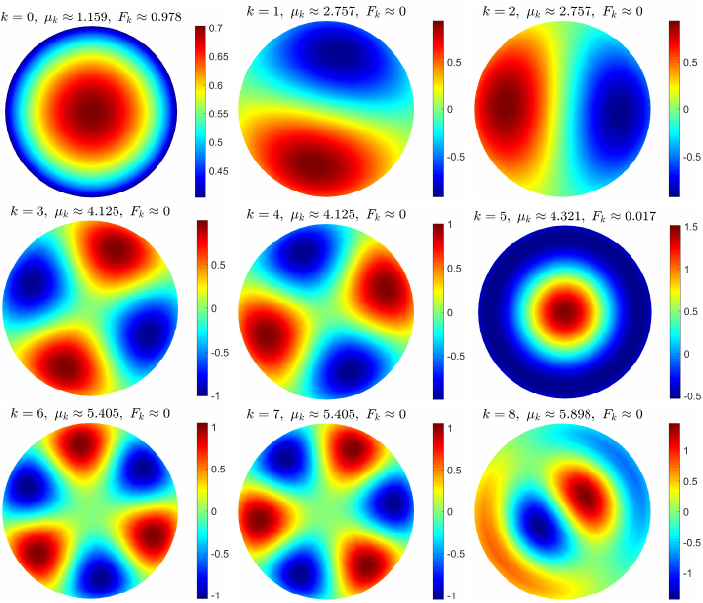} % {disk_Vk1.pdf}
\end{center}
\caption{
First 9 Steklov eigenfunctions $\Psi_k$ restricted to the circular
patch of unit radius (the associated eigenvalues $\mu_k$ and weights
$F_k$ are shown in the titles).  These eigenfunctions were obtained by
using a mesh with 688 triangles and 375 nodes. }
\label{fig:disk_Vk}
% [Ed,V,dk, p,e,t,Gam,Geom] = A_DN_patch_FEM_ellipses(1, 0);
%  A_DN_patch_FEM_show(V,1,Ed, p,e,t);
%  'disk.mat',  'diskN.mat'
\end{figure}

Figure \ref{fig:disk_Vk} presents the first 9 Steklov eigenfunctions
$\Psi_k|_{\Gamma}$ on the circular patch of unit radius.  The
associated eigenvalues $\mu_k$ and weights $F_k$ are shown in the
panels.  As discussed in Appendix \ref{sec:validation} (see also
\cite{Grebenkov24}), the axial symmetry of the problem implies that
the eigenfunctions can be written in polar coordinates $(r,\phi)$ as
$e^{\pm im\phi} v_{m,n}(r)$, with positive integer $m$ and $n$.  For
any $m \ne 0$, a linear combination of two non-axially symmetric
eigenfunctions $e^{im\phi} v_{m,n}(r)$ and $e^{-im\phi} v_{m,n}(r)$ is
also an eigenfunction, implying that the associated eigenvalue is at
least twice degenerate.  This is confirmed by the shape of the shown
eigenfunctions (which are chosen to be real-valued), e.g., $\Psi_1$
and $\Psi_2$ correspond to the same eigenvalue.  Our numerical results
suggest that the eigenvalues corresponding to axially symmetric
eigenfunctions are simple, whereas the eigenvalues corresponding to
non-axially symmetric eigenfunctions are twice degenerate.  We stress
that only axially symmetric eigenfunctions (here, with indices $k=0$
and $k = 5$) do contribute to the reactive capacitance $C(\mu)$
because the projection of any non-axially symmetric eigenfunction onto
a constant is zero.

\subsubsection*{Elliptic patches}

The trapping capacity of a perfectly reactive elliptic patch was
studied by Strieder \cite{Strieder09}.  In particular, the exact
solution of Eq. (\ref{eq:wi_def}) at $\mu = \infty$, that determines
the concentration profile $w(\y;\infty)$ in the upper half-space, was
given in ellipsoidal coordinates (see also \cite{Morse}), from which
the flux density in Eq. (\ref{eq:q_elliptic}) follows.  Integrating
this density over the patch, one retrieves the electrostatic
capacitance \cite{Landkof}:
% see examples on pages 165-167. 
%
\begin{equation}  \label{eq:Cinf_ellipse}
C(\infty) =  \frac{b}{K(\sqrt{1 - a^2/b^2})} \,.
\end{equation}
Unfortunately, these analytical results do not admit explicit
extensions to partially reactive elliptic patches.  In turn, one can
express the flux density and the reactive capacitance in terms of the
Steklov eigenfunctions.

\begin{figure}
\begin{center}
\includegraphics[width=\columnwidth]{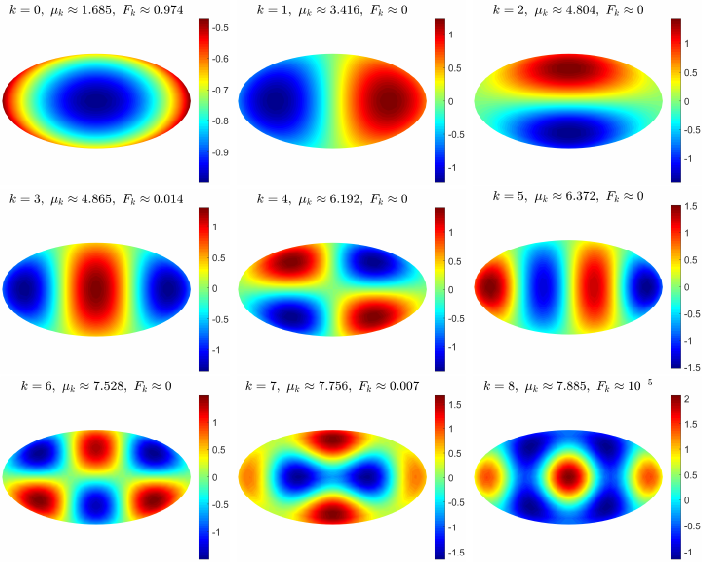} % {ellipse_Vk1.pdf}
\end{center}
\caption{
First 9 Steklov eigenfunctions $\Psi_k$ restricted to the elliptic
patch of semiaxes $0.5$ and $1$ (the associated eigenvalues $\mu_k$
and weights $F_k$ are shown in the titles).  The eigenfunctions were
obtained by using a mesh with 756 triangles and 413 nodes. }
\label{fig:ellipse_Vk}
% [Ed,V,dk, p,e,t,Gam,Geom] = A_DN_patch_FEM_ellipses(0.5, 0);
%  A_DN_patch_FEM_show(V,1,Ed, p,e,t);
%  'ellipse05.mat'
\end{figure}

Figure \ref{fig:ellipse_Vk} presents the first 9 eigenfunctions
$\Psi_k|_\Gamma$ and related eigenvalues $\mu_k$ for an elliptic patch
with semiaxes $1$ and $0.5$.  In contrast to the circular patch, all
shown eigenvalues are simple, i.e., anisotropy generally removes the
degeneracy of eigenvalues.  At the same time, the continuous
dependence of the eigenvalues on the aspect ratio $a/b$ allows one to
construct ellipses, for which some eigenvalues are degenerate (e.g.,
the shown eigenvalues $\mu_2$ and $\mu_3$ are close, and they can be
made equal by varying $a/b$).  The symmetry of the elliptic patch
implies that some eigenfunctions are antisymmetric with respect to the
horizontal or vertical axis, so that their projection onto a constant
is zero.  Such antisymmetric eigenfunctions do not contribute to the
reactive capacitance.  Among the shown eigenfunctions, only $\Psi_0$,
$\Psi_3$, $\Psi_7$ and $\Psi_8$ do contribute to $C(\mu)$, with their
relative weights: $F_0 \approx 0.9741$, $F_3 \approx 0.0138$, $F_7
\approx 0.0074$ and $F_8 \approx 10^{-5}$.  One sees that, as for the
circular patch, the first eigenfunction $\Psi_0$ provides the dominant
contribution of $97\%$.

\begin{figure}
\begin{center}
\includegraphics[width=\columnwidth]{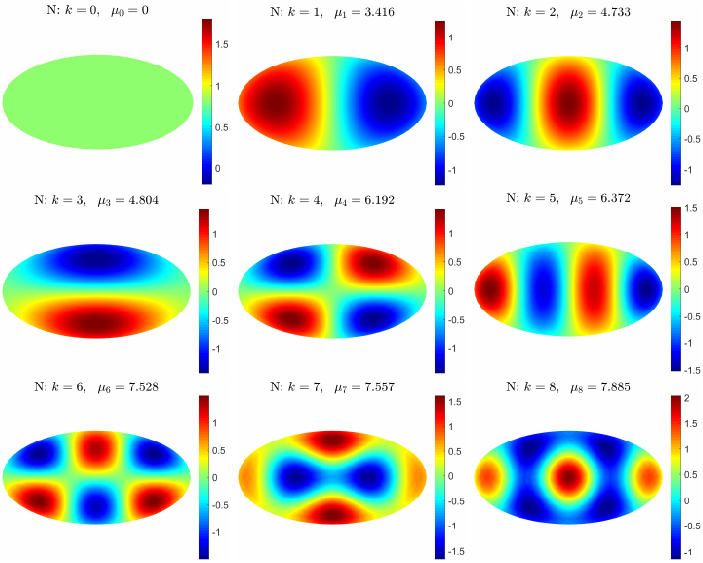} % {ellipseN_Vk1.pdf}
\end{center}
\caption{
First 9 Steklov eigenfunctions $\Psi_k^N$ restricted to the elliptic
patch of semiaxes $0.5$ and $1$ for the Neumann version (the
associated eigenvalues are shown in the titles).  The eigenfunctions
were obtained by using a mesh with 756 triangles and 413 nodes. }
\label{fig:ellipseN_Vk}
% [Ed,V,dk, p,e,t,Gam,Geom] = A_DN_patch_FEM_ellipses(0.5, 1);
%  A_DN_patch_FEM_show(V,1,Ed, p,e,t);
%  'ellipse05N.mat'
\end{figure}

We complete this discussion by illustrating the difference between the
Steklov problems (\ref{eq:Psi_def}) and (\ref{eq:PsiN_def}).  Figure
\ref{fig:ellipseN_Vk} presents the first 9 eigenfunctions
$\Psi_k^N|_\Gamma$ for the elliptic patch of semiaxes $1$ and $0.5$.
Their comparison with panels of Fig. \ref{fig:ellipse_Vk} indicates
that, apart from the principal eigenfunctions $\Psi_0$ and $\Psi_0^N$,
which are different, the other eigenfunctions look quite similar, up
to matching their indices.  Moreover, antisymmetric eigenfunctions are
actually identical, up to their signs, e.g. $\Psi_1^N = -\Psi_1$,
$\Psi_3^N = -\Psi_2$, etc.  This is not surprising (see Appendix
\ref{sec:FEM} for details): as the kernels $\G(\y,\y')$ and
$\G^N(\y,\y')$ differ by a function $\omega_\Gamma(\y)/|\Gamma| +
\omega_\Gamma(\y')/|\Gamma| - \A_\Gamma$, which is symmetric on the
elliptic patch, the antisymmetric eigenfunctions of these kernels
should be identical.

\subsubsection*{Rectangular patches}

\begin{figure}
\begin{center}
\includegraphics[width=\columnwidth]{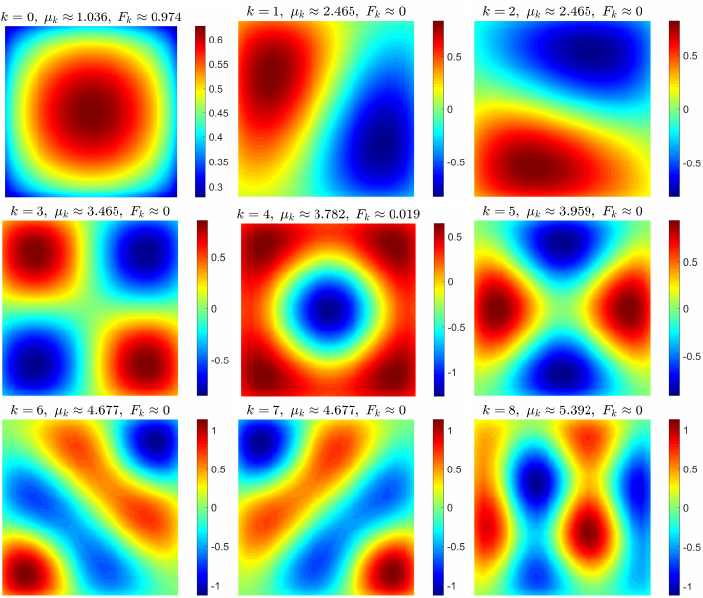} % {square_Vk1.pdf}
\end{center}
\caption{
First 9 Steklov eigenfunctions $\Psi_k$ restricted to the square patch
$(-1,1)\times (-1,1)$ (the associated eigenvalues $\mu_k$ and weights
$F_k$ are shown in the titles).  The eigenfunctions were obtained by
using a mesh with 674 triangles and 372 nodes. }
\label{fig:square_Vk}
% [Ed,V,dk, p,e,t,Gam,Geom] = A_DN_patch_FEM_rectangles(1, 0);
%  A_DN_patch_FEM_show(V,1,Ed, p,e,t);
%  See 'square.mat'
\end{figure}

\begin{figure}
\begin{center}
\includegraphics[width=\columnwidth]{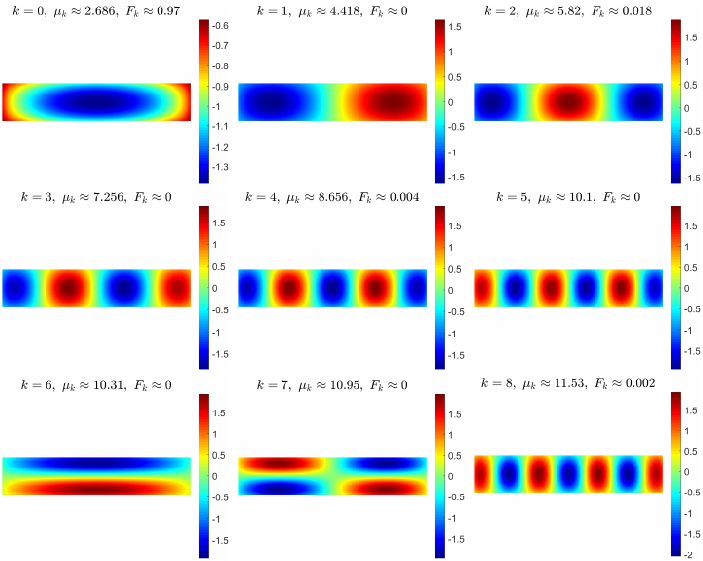} % rectangle_Vk1.pdf}
\end{center}
\caption{
First 9 Steklov eigenfunctions $\Psi_k$ restricted to the rectangular
patch $(-1,1)\times (-0.2,0.2)$ (the associated eigenvalues $\mu_k$
and weights $F_k$ are shown in the titles).  The eigenfunctions were
obtained by using a mesh with 704 triangles and 400 nodes. }
\label{fig:rectangle02_Vk}
% [Ed,V,dk, p,e,t,Gam,Geom] = A_DN_patch_FEM_rectangles(0.2, 0);
%  A_DN_patch_FEM_show(V,1,Ed, p,e,t);
%  See 'rectangle_02.mat'
\end{figure}

Figures \ref{fig:square_Vk} and \ref{fig:rectangle02_Vk} present
similar results for a quadratic patch $(-1,1)\times (-1,1)$ and for an
elongated rectangular patch $(-1,1)\times (-0.2,0.2)$.  As in the case
of a circular patch, symmetries of the square make some eigenvalues
degenerate, e.g., $\mu_1 = \mu_2$ and $\mu_6 = \mu_7$.  In turn, this
degeneracy is generally removed for rectangular patches, even though
continuous variations of $a/b$ can be used to obtained some degenerate
eigenvalues.

\subsubsection*{Disconnected patches}

To give a broader view onto the spectral properties, we also discuss
the case of disconnected patches.  In sharp contrast to Laplacian
eigenfunctions, for which the connectivity of the domain plays the
crucial role, it is less relevant for the considered Steklov problem.
For instance, splitting a patch into two disconnected subsets by a
curve has no effect onto the spectrum.  Moreover, since the kernel
$\G(\y,\y')$ decays slowly with the distance between $\y$ and $\y'$,
even if two subsets are separated by a significant distance, they
still compete with each other for capturing diffusing particles.  This
long-range interaction, known as diffusional screening
\cite{Sapoval94,Sapoval02,Felici03,Felici05} or diffusive interaction
\cite{Traytak92}, implies that the electrostatic capacitance is not
additive, i.e., the capacitance of the union of two patches is always
smaller than the sum of their individual capacitances.  For perfect
reactions, this competition for the case of two circular patches on
the reflecting plane was characterized analytically in
\cite{Strieder08,Saddawi12}, whereas Lindsay and Bernoff derived
asymptotic formulas in the case of multiple small circular patches
\cite{Lindsay18} (see also \cite{Bernoff18}).  A further extension to
partially reactive patches of arbitrary shape was given in
\cite{GrebWard25b}.  It is therefore instructive to inspect the
behavior of the Steklov eigenfunctions in the case of disconnected
patches.

\begin{figure}
\begin{center}
\includegraphics[width=\columnwidth]{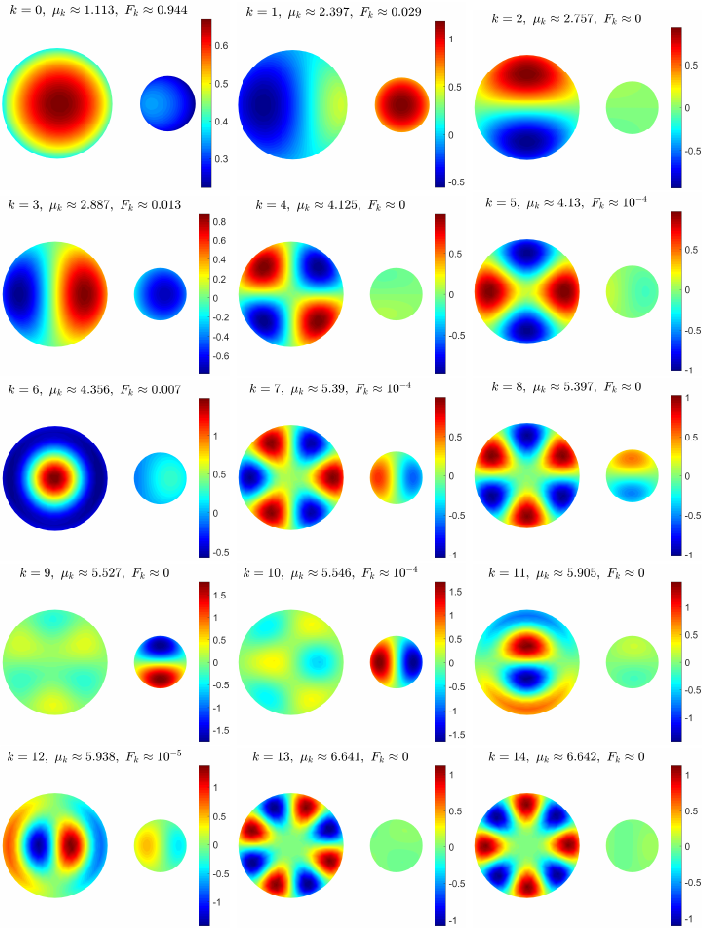} % twodisk_Vk1.pdf}
\end{center}
\caption{
First 15 Steklov eigenfunctions $\Psi_k$ restricted to the patch
formed by two disks of radii $1$ and $0.5$, separated by distance
$0.5$ (the associated eigenvalues $\mu_k$ and weights $F_k$ are shown
in the titles).  The eigenfunctions were obtained by using a mesh with
964 triangles and 536 nodes. }
\label{fig:twodisk_Vk}
% [Ed,V,dk, p,e,t,Gam,Geom] = A_DN_patch_FEM_ellipses(1, 0);
%%   One needs to uncomment [p,e,t] = A_DN_patch_FEM_replicate_patch(p,e,t, [2 0 0.5]');
%%   One needs to uncomment [p,e,t] = A_DN_patch_FEM_replicate_patch(p,e,t, [3 0 1.5]');
%  A_DN_patch_FEM_show(V,1,Ed, p,e,t);
\end{figure}

Figure \ref{fig:twodisk_Vk} shows the first 15 Steklov eigenfunctions
$\Psi_k|_\Gamma$ for two circular patches of radii $1$ and $0.5$,
separated by distance $0.5$.  Interestingly, the presence of the
smaller patch has only a minor effect on the principal eigenvalue
$\mu_0$ and the spectral weight $F_0$ of the associated eigenfunction.
We also note that, despite the proximity of two patches, some
eigenfunctions are localized on one patch (i.e., they almost vanish on
the other).  The refined structure of the Steklov eigenpairs gives
access to the reactive capacitance and its dependence on the spatial
arrangement of the disconnected patches.  Further analysis of this
problem presents an interesting perspective of this work.

\subsection{Effect of patch anisotropy onto the eigenvalues}

To investigate more quantitatively the role of patch anisotropy, we
inspect three families of patches: (i) ellipses of semi-axes $a$ and
$b$, (ii) rectangles $(-a,a)\times (-b,b)$, and (iii) rhombuses with
diagonals $2a$ and $2b$.  The patch surface areas are respectively
$\pi ab$, $4ab$, and $2ab$.  We fix $b = 1$ and then change the
anisotropy by varying $a$ from $0$ to $1$.  In the limit $a \to 0$,
all three shapes approach the interval $(-1,1)$, which is inaccessible
to three-dimensional Brownian motion and thus presents a singular
limit.

\begin{figure}
\begin{center}
\includegraphics[width=0.99\columnwidth]{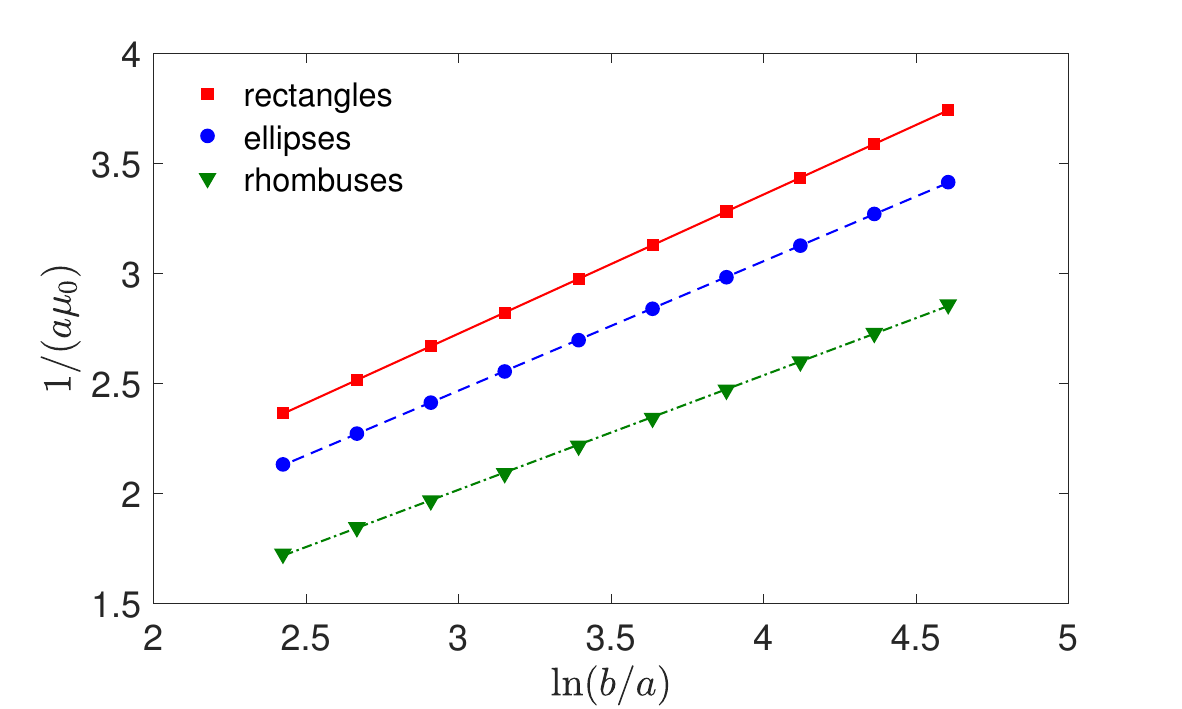} % {mu0new.eps}
\end{center}
\caption{
Asymptotic behavior of the principal eigenvalue $\mu_0$ for elliptic,
rectangular and rhombic patches with $b = 1$ and variable $a$.
Symbols show $1/(a\mu_0)$, obtained numerically, as a function of
$\ln(b/a)$, whereas lines indicate linear fits given in the text.}
\label{fig:mu0}
% [mu1,mu2,mu3,F1,F2,F3,a] = A_DN_patch_FEM_mu_fig();
%%%% A_Raphael_DN_load_muk_fig(mu, d, a, mu0, d0, a0);
\end{figure}

We first inspect the asymptotic behavior of the eigenvalues $\mu_k$
for an elliptic patch as $a\to 0$.  Since the elliptic patch shrinks
to an interval, the eigenvalues are expected to diverge as $a
\to 0$.  The principal eigenvalue $\mu_0$ cannot diverge faster than
the right-hand side of the Payne's inequality (\ref{eq:mu0_upper}).
Substituting the asymptotic behavior 
\begin{equation}  \label{eq:K_asympt}
K(z) \sim 2\ln(2) - \frac12 \ln(1 - z^2) \qquad (z\to 1)
\end{equation}
to Eq. (\ref{eq:Cinf_ellipse}), one gets
\begin{equation}  \label{eq:Cinf_ellipses}
C(\infty) \sim b/\ln(4b/a)  \qquad (a\to 0).
\end{equation}
As a consequence, $\mu_0$ cannot grow faster than $\propto (a \ln
(4b/a))^{-1}$.  Figure \ref{fig:mu0} shows $1/(a\mu_0)$ (circles) as a
function of $\ln(b/a)$ to confirm this behavior.  Indeed, one observes
a linear dependence,
\begin{equation}  \label{eq:mu0_ellipse_fit}
\frac{1}{a\mu_0} \approx 0.59 \ln(b/a) + 0.71  \qquad (a\ll b),
\end{equation}
with the coefficients obtained from a linear fit (shown by dashed
line).

To our knowledge, there is no analytic expression for the capacitance
of a rectangular patch.  For this reason, we employ the asymptotic
behavior (\ref{eq:A_rectangle_asympt}) of the coefficient $\A_\Gamma$
for elongated rectangles (see Appendix \ref{sec:wGamma}).  According
to the lower and upper bounds (\ref{eq:mu0_boundsA}), the asymptotic
behavior of $\A_\Gamma$ is expected to control that of the principal
eigenvalue $\mu_0$.  Substituting Eq. (\ref{eq:A_rectangle_asympt})
into Eq. (\ref{eq:mu0_boundsA}), we get approximately
\begin{equation}  \label{eq:mu0_rectangle_asympt}
\mu_0 \approx \frac{1}{|\Gamma| \A_\Gamma} 
\simeq \frac{\pi}{a(1 + 2\ln (2b/a))}  \qquad (a\ll b),
\end{equation}
or, equivalently,
\begin{equation}  \label{eq:mu0_rectangle_fit}
\frac{1}{a\mu_0} \approx \frac{1 + 2\ln (2b/a)}{\pi} \approx 0.64 \ln (b/a) + 0.76 \qquad (a\ll b).
\end{equation}
Figure \ref{fig:mu0} presents by squares the left-hand side of this
relation, with $\mu_0$ obtained numerically.  A linear fit of these
values is shown by solid line and yields $1/(a\mu_0) \approx 0.63
\ln(b/a) + 0.83$, in an excellent agreement with our
prediction (\ref{eq:mu0_rectangle_fit}).

A similar asymptotic analysis could be performed for rhombic patches.
However, we skip this analysis and just present a linear fit of
$1/(a\mu_0)$ versus $\ln(b/a)$, obtained from numerical values of
$\mu_0$ and shown in Fig. \ref{fig:mu0} by dash-dotted line:
\begin{equation}  \label{eq:mu0_rhombuses_fit}
\frac{1}{a\mu_0} \approx 0.52 \ln(b/a) + 0.45   \qquad (a\ll b).
\end{equation}

In all three cases, the asymptotic behavior is similar but the
coefficients are slightly different.  We outline that the numerical
results in Fig. \ref{fig:mu0} are in agreement with the domain
monotonicity property (\ref{eq:muk_monotonicity}), i.e., the curves of
$1/(a\mu_0)$ are ordered according to the fact that a rhombus is
inscribed into an ellipse, which in turn is inscribed into a rectangle
(of the same aspect ratio $a/b$).  In particular, the numerical
coefficients in front of $\ln(b/a)$ are $0.52$, $0.59$ and $0.64$ for
these three shapes.

For all three shapes, we also observed the asymptotic behavior $\mu_k
\propto 1/(a\ln(b/a))$ for some other eigenvalues (not shown).  
It is worth noting that a similar behavior of the principal eigenvalue
was obtained for the exterior Steklov problem for elongated prolate
spheroids with semiaxes $a$, $a$, $b$ \cite{Grebenkov24}.  The
explicit construction of a matrix representation of the associated
Dirichlet-to-Neumann operator allowed one to show that $\mu_{0,n}
\propto 1/(a\ln(b/a))$ as $a\to 0$ for the eigenvalues $\mu_{0,n}$
corresponding to axially symmetric eigenfunctions.

\subsection{Spectral weights}

\begin{figure}[h!]
\begin{center}
\includegraphics[width=0.99\columnwidth]{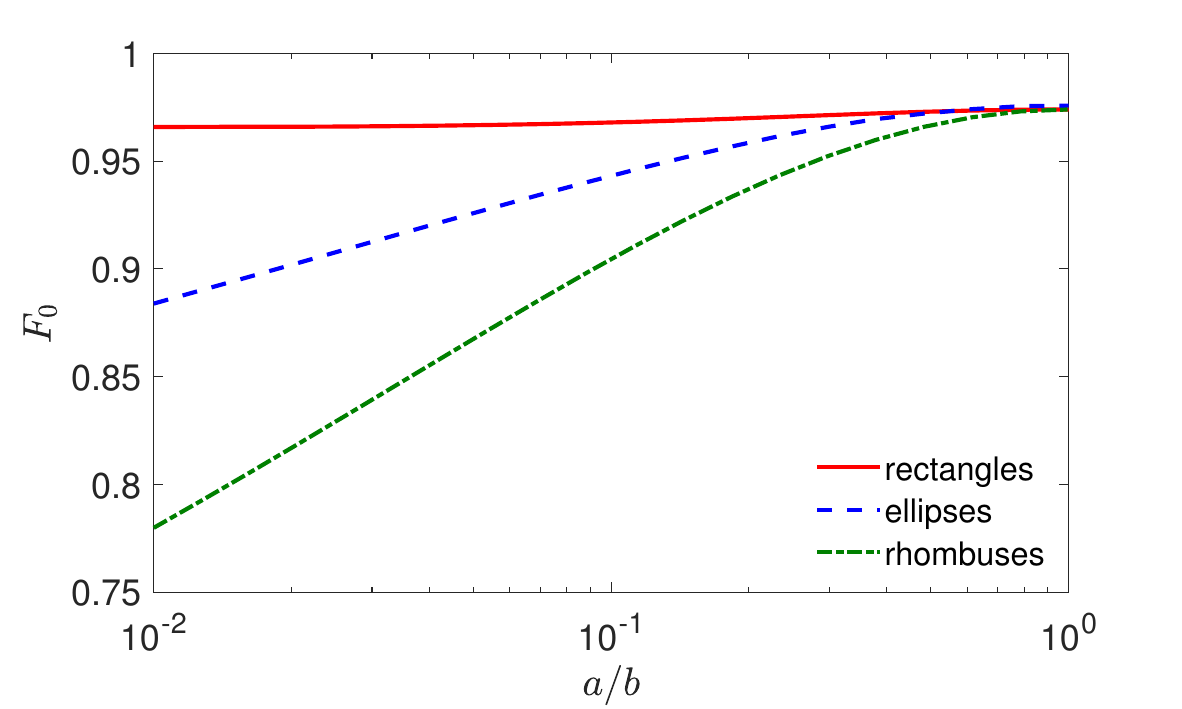} % F0new.eps}
\end{center}
\caption{
Relative contribution $F_0$ of the principal eigenmode, obtained
numerically, as a function of $a/b$, for rectangles, ellipses and
rhombuses, with $b = 1$ and variable $a$.  }
\label{fig:F0}
% [mu1,mu2,mu3,F1,F2,F3,a] = A_DN_patch_FEM_Fk_fig();
%%%%% [mu,d,a, mu0,d0,a0, mu2,d2,a2] = A_Raphael_DN_load_dk_fig;
\end{figure}

Next, we look at the coefficient $F_0$ that represents the relative
contribution of the principal eigenmode to the reactive capacitance.
For a circular patch, one has $F_0 \approx 0.97$ \cite{GrebWard25a},
i.e., the principal eigenmode provides the dominant contribution.  A
similar observation was reported for the exterior Steklov problem for
tori and two balls \cite{Grebenkov25a}.  We inspect how the patch
anisotropy may affect the coefficient $F_0$.  Figure
\ref{fig:F0} shows $F_0$ as a function of the aspect ratio $a/b$ for
three considered shapes.  The first observation is that $F_0$ remains
the dominant contribution for all three shapes on the considered range
$0.01 \leq a/b \leq 1$.  Curiously, the behavior is quite different
between rectangles and elliptic/rhombic patches: the coefficient $F_0$
is almost constant for rectangles, whereas it exhibits a very slow
decay for ellipses and rhombuses as $a$ decreases.  A clarification of
this distinction remains an interesting open problem.

The empirical observation that $F_0$ remains above $0.78$ for the
considered domains and aspect ratios raises several geometric
questions.  Even though $F_0$ decreases on the range $0.01 \leq a/b
\leq 1$, it is yet unclear whether this decay holds in the limit
$a\to 0$.  Moreover, even if $F_0 \to 0$ as $a\to 0$, such a slow
decay requires to consider extremely thin patches to achieve small
$F_0$.  Is it possible to construct other types of shapes to diminish
$F_0$?  While the actual value of $F_0$ does not look relevant, the
closeness of $F_0$ to $1$ may have important consequences.  For
instance, the upper and lower bounds (\ref{eq:mu0_upper_cn},
\ref{eq:mu0_lower_cn}) are close when $F_0 \approx 1$, yielding thus
an accurate estimate for the principal eigenvalue $\mu_0$.  Moreover,
we showed in Sec. \ref{sec:approx} that the accuracy of the sigmoidal
approximation (\ref{eq:Cmu_approx}) is higher when $F_0$ is close to
$1$.  For these reasons, further analysis and eventual decrease of
$F_0$ may provide a better understanding of the reactive capacitance.

For such an attempt, we consider a rectangular dumbbell patch formed
by two squares that are connected by a long thin rectangular channel.
When the channel width goes to $0$, the connection progressively
disappears, and the dumbbell patch reduces to two disconnected
squares.  The asymptotic behavior of the eigenvalues and
eigenfunctions of the Laplace operator in dumbbell domains was
thoroughly investigated in the past (see \cite{Grebenkov13} and
references therein).  In particular, Laplacian eigenfunctions can be
localized in one part of the dumbbell when they cannot ``squeeze''
through a thin channel, and this behavior strongly depends on the type
of the boundary condition (Dirichlet vs Neumann).  In contrast, the
connectivity of a patch is expected to be much less relevant for the
exterior Steklov problem.  In fact, even in the disconnected setting,
there are long-range interactions between distinct parts of the patch
via the kernel $\G(\y,\y') = 1/(2\pi |\y-\y'|)$.

\begin{figure}
\begin{center}
\includegraphics[width=\columnwidth]{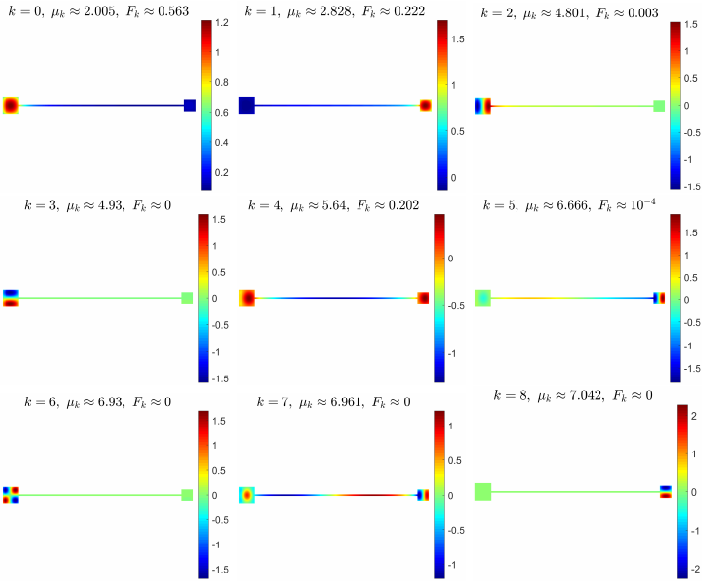} % {dumbbell_Vk1.pdf}
\end{center}
\caption{
First 9 Steklov eigenfunctions $\Psi_k$ on a rectangular dumbbell
patch formed by two squares of lengthsides $1$ and $0.7$ and connected
by a rectangular channel of length $10$ and width $0.1$ (the
associated eigenvalues $\mu_k$ and weights $F_k$ are shown in the
titles).  The eigenfunctions were obtained by using a mesh with 1499
triangles and 965 nodes. }
\label{fig:dumbbell_Vk}
% [Ed,V,dk,Fk, p,e,t,Gam,Geom] = A_DN_patch_FEM_dumbbell();
%  A_DN_patch_FEM_show(V,1,Ed, p,e,t);
%  See 'dumbbell.mat'
\end{figure}

Figure \ref{fig:dumbbell_Vk} illustrates the first 9 eigenfunctions
$\Psi_k|_\Gamma$ on such a dumbbell patch.  The first and most
important observation is that the weight $F_0 \approx 0.56$ of the
principal eigenfunction is diminished as compared to that of elliptic,
rectangular and rhombic patches.  In turn, the contribution of the
eigenfunctions $\Psi_1$ and $\Psi_4$ became more significant, with
$F_1 \approx 0.22$ and $F_4 \approx 0.20$.  Interestingly, we also
observe the effect of localization of some eigenfunctions in either of
two squares, e.g., $\Psi_2$, $\Psi_3$ and $\Psi_6$ are localized in
the larger square, whereas $\Psi_8$ is localized in the smaller
square.  Note that a minor shift of the smaller square breaks the
mirror symmetry of the considered dumbbell patch with respect to the
horizontal axis, but does not affect the localization of
eigenfunctions (not shown).

\subsection{Sigmoidal approximation}

We inspect the quality of the explicit approximation $C^{\rm
app}(\mu)$ from Eq.  (\ref{eq:Cmu_approx}).  For this purpose, we
evaluate the maximal relative error of $C^{\rm app}(\mu)$ as compared
to $C(\mu)$, over a broad range of $\mu$ from $10^{-2}$ to $10^2$.
Since the approximation (\ref{eq:Cmu_approx}) reproduces correctly the
leading-order asymptotic behavior of $C(\mu)$ in both limits $\mu\to
0$ and $\mu\to \infty$, this estimate is expected to actually cover
the whole range of $\mu$ from $0$ to $\infty$.  For the considered
family of elliptic patches with $a$ from $0.01$ to $1$, the maximal
relative error takes the largest value of $4.1\%$ for the circular
patch ($a = 1$), and then decreases for smaller $a$.  A similar trend
was observed for rectangular patches, with the maximal relative error
of $4.4\%$ corresponding to the quadratic patch.
%  [dC,a] = A_DN_patch_FEM_Cmu_check();

According to Eq. (\ref{eq:Capprox_error}), the relative error of the
sigmoidal approximation cannot exceed $\Emax$.  In the case of
elliptic patches, one can use the exact relations
(\ref{eq:Cinf_ellipse}, \ref{eq:AGamma_ellipse}) for $C(\infty)$ and
$\A_\Gamma$ to get $\Emax = 32/(3\pi^2) - 1 \approx 8.1\%$.  This
conservative bound is twice larger than the actual maximal relative
error.  For the unit square patch, Eq. (\ref{eq:AGamma_square}) yields
$\A_\Gamma \approx 0.4733$, whereas the electrostatic capacitance was
found numerically to be $C(\infty) \approx 0.3667874$ \cite{Read97} so
that $\Emax \approx 9.1\%$.  This bound is again twice larger than the
actual maximal relative error.

\section{Discussion and conclusion}
\label{sec:conclusion}

In this paper, we studied the reactive capacitance $C(\mu)$ for planar
patches of arbitrary shape.  The spectral representation
(\ref{eq:Cmu}) determines its functional dependence on the reactivity
parameter $\mu$, in which the geometric properties of the patch are
incorporated through the eigenvalues $\mu_k$ and the weights $F_k$.
Moreover, we developed an efficient numerical tool for computing these
geometric parameters for a given patch.  This tool was applied to
inspect the spectral properties and the resulting reactive
capacitances of various patches.  We also derived variational
formulations, probabilistic interpretations and various bounds for the
reactive capacitance.  Apart from these theoretical achievements, we
obtained several practical results with potential applications in
chemical physics that we summarize below.

The first practical result is that the principal eigenmode provides
the dominant contribution to the reactive capacitance.  On one hand,
an important role of the principal eigenmode could be expected;
indeed, as $\Psi_0$ is the unique eigenfunction that does not change
sign on the patch, its projection onto a constant cannot vanish, i.e.,
$F_0 > 0$.  In turn, the other eigenfunctions $\Psi_k$ are orthogonal
to $\Psi_0$ and thus must change sign, so that their projections onto
a constant are expected to be smaller.  On the other hand, our
numerical results indicate that the contribution $F_0$ exceeds $96\%$
for all considered rectangular patches, irrespective of their aspect
ratio (see Fig. \ref{fig:F0}).  For elliptic and rhombic patches, the
situation is different: the relative weight $F_0$ is still dominant
but it decreases very slowly as the aspect ratio $a/b$ goes to $0$.
However, even for the value $a/b$ as small as $0.01$, the principal
eigenmode provides $88\%$ for the elliptic patch and $78\%$ for the
rhombic patch.  The dominant role of $F_0$ suggests that the
associated principal eigenvalue determines the most relevant
lengthscale of the problem as $1/\mu_0$.  Rewriting
Eq. (\ref{eq:Fkmuk_sum}) as
\begin{equation}
C(\infty) = \frac{|\Gamma|}{2\pi} \biggl( \mu_0 \underbrace{\sum\limits_{k=0}^\infty F_k}_{=1} 
+ \sum\limits_{k=1}^\infty F_k(\mu_k - \mu_0)\biggr),
\end{equation}
and neglecting the second sum, one gets a rough approximation $\mu_0
\approx 2\pi C(\infty)/|\Gamma|$ (in practice, the right-hand side is
the Payne's upper bound (\ref{eq:mu0_upper})).  For instance, for the
circular patch of unit radius, this approximation gives $\mu_0
\approx 4/\pi \approx 1.27$, whereas the actual numerical value is
$1.159$, i.e., one gets only $10\%$ relative error.  More accurate
bounds that involve the constant $\A_\Gamma$ were discussed in
Sec. \ref{sec:bounds}.  A systematic analysis of the accuracy of these
approximations presents an interesting perspective.

Another related open problem is to characterize the patches, for which
$F_0$ is smaller than a given threshold.  According to our numerical
results, one may probably achieve this condition by taking extremely
thin elliptic or rhombic patches, but are there other ways to diminish
$F_0$?  We provided an example of a dumbbell domain, for which the
relatively low value $F_0 \approx 0.56$ could result from the
localization phenomenon.  However, a rigorous analysis of this
phenomenon via local estimates on the eigenfunctions seems to be much
more challenging than in the conventional setting of Laplacian
eigenfunctions (see \cite{Grebenkov13}).
Moreover, one can inspect the contribution of other eigenfunctions and
ask whether there exist patches, for which the relative contribution
of the principal eigenfunction is not maximal?  Even though these
questions sound rather abstract, a better understanding of the
relation between the shape of the patch and the geometry of its
Steklov eigenfunctions may help to design more efficient catalysts
with desired reactive properties.

The second practically important result is that the simple sigmoidal
approximation $C^{\rm app}(\mu)$ from Eq. (\ref{eq:Cmu_approx0}) turns
out to be accurate for a variety of patch shapes that we considered.
This numerical observation suggests that, for many practical purposes,
it is sufficient to know the electrostatic capacitance $C(\infty)$ and
the surface area $|\Gamma|$ of the patch.  The additive form of
Eq. (\ref{eq:Cmu_approx0}) allows one to interpret $C^{\rm app}(\mu)$
as a consecutive connection of a ``diffusion resistance''
$1/C(\infty)$ and a ``reaction resistance'' $2\pi/(\mu |\Gamma|)$, in
the language of electrostatics.  In fact, as a resistance is the ratio
between the applied voltage and the induced electric current,
$1/C(\mu)$ plays the role of a resistance in diffusion-reaction
problems.  The above interpretation was proposed as an empirical way
to describe the Laplacian transport, which is decomposed into the
first diffusive step and the second reactive step
\cite{Sapoval94,Sapoval02,Felici03,Felici05}.  Its mathematical
validation for electrochemical transport was provided in
\cite{Grebenkov06b}, while our work justifies it for steady-state
diffusion-controlled reactions on partially reactive flat patches.
Moreover, we obtained an upper bound (\ref{eq:Capprox_error}) for the
relative error of the sigmoidal approximation, which is expressed in
terms of $|\Gamma|$, $C(\infty)$, and the constant $\A_\Gamma$ defined
in Eq. (\ref{eq:A_def}).  We derived a useful representation
(\ref{eq:AGamma_general}) for an efficient numerical computation of
$\A_\Gamma$ for general polygonal patches and provided the exact
explicit formulas for elliptic and rectangular patches.  The versatile
roles of the constant $\A_\Gamma$ in various contexts were discussed
in \cite{Grebenkov25a}.  

The reactive capacitance was shown to determine the small-patch
asymptotic behavior of many characteristics of diffusion-controlled
reactions, such as the mean first-passage time, splitting
probabilities, and the total flux \cite{GrebWard25a,GrebWard25b}.  For
instance, for restricted diffusion inside a bounded domain $\Omega$
with a reflecting boundary covered by multiple small reactive patches,
well-separated from each other, the leading-order term of the
volume-averaged mean first-reaction time is
\begin{equation}
T \sim \frac{|\Omega|}{2\pi D \overline{C}} \,,
\end{equation}
where $\overline{C}$ is the sum of reactive capacitances of the
patches, $D$ is the diffusivity, and $|\Omega|$ is the volume of the
domain.  One sees that the reactive capacitance plays the role of an
effective patch size that incorporates its shape and reactivity.  For
a single patch, our sigmoidal approximation (\ref{eq:Cmu_approx0})
yields
\begin{equation}
T \sim |\Omega| \biggl(\frac{1}{2\pi D C(\infty)} + \frac{1}{\kappa |\Gamma|}\biggr),
\end{equation}
in agreement with the above discussion on the consecutive connection
of the diffusive and reactive steps.  Two next-order correction terms
were derived in \cite{GrebWard25a}. 

Following \cite{Chaigneau22,Grebenkov25a}, we also introduce the {\it
reactive length} $L_\Gamma = |\Gamma|/(2\pi C(\infty))$, associated to
a flat patch of arbitrary shape, in order to rewrite the approximate
reactive capacitance as $C^{\rm app}(\mu) = C(\infty)/(1 + 1/(\mu
L_{\Gamma}))$.  This expression indicates how the reactivity of the
patch reduces its electrostatic capacitance $C(\infty)$, and
$L_{\Gamma}$ is the proper geometric lengthscale that is compared to
the physical lengthscale $1/\mu = D/\kappa$ (see
\cite{Sapoval94,Sapoval02} for further discussions).  For instance,
the reactive length of a circular patch of radius $a$ is
$\tfrac{\pi}{4} a$, whereas the reactive length of a quadratic patch
with edge length $L$ is $L_{\Gamma} \approx 0.43\, L$ (here we used
the numerical value $C(\infty) \approx 0.3667874$ for the capacitance
of the unit square \cite{Read97}).  For an elliptic patch with
semiaxes $a < b$, we get
\begin{equation}
L_\Gamma = a \frac{K(\sqrt{1-a^2/b^2})}{2}  \,.
\end{equation}
When the patch is elongated ($a\ll b$), one finds $L_\Gamma
\approx \tfrac{1}{2} a \ln(4b/a)$, i.e., the reactive length vanishes
as $a\to 0$, as expected.

While this paper was focused on flat patches, most results can be
immediately generalized to bounded targets of arbitrary shape with
sufficiently smooth boundary in $\R^3$.  In turn, the large-$\mu$
behavior of $C(\mu)$ that originated from the edge singularity, as
well as the domain monotonicity of the Steklov eigenvalues, are not
valid in general.  The efficient numerical tool that we developed here
is also not applicable for general targets.  In some special cases,
the symmetries of the target allow one to employ orthogonal
curvilinear coordinates to represent Steklov eigenfunctions on the
basis of special functions.  This technique was used for prolate and
oblate spheroids in \cite{Grebenkov24}, as well as for a pair of balls
and a torus in \cite{Grebenkov25a}.  For instance, the spectral
expansion (\ref{eq:Cmu}), the role of the principal weight $F_0$, and
the accuracy of the sigmoidal approximation (\ref{eq:Cmu_approx}) were
discussed for these targets in \cite{Grebenkov25a}.  However, the
numerical computation of Steklov eigenmodes for an arbitrarily-shaped
target requires the knowledge of the Green's function $G(\y,\y')$ in
the exterior of the target, as well as a suitable boundary integral
method.  A further development in this direction presents an
interesting perspective.  More generally, the proposed theoretical
description of the reactive capacitance opens promising ways for
modeling diffusion-reaction phenomena in complex media and for solving
shape-optimization problems in chemical engineering and other
disciplines.

\begin{acknowledgments}
The authors thank Prof. N. Nigam and Prof. I. Polterovich for fruitful
discussions.  D. S. G. acknowledges the Simons Foundation for
supporting his sabbatical sojourn in 2024 at the CRM, University of
Montr\'eal, Canada, as well as the Alexander von Humboldt Foundation
for support within a Bessel Prize award.
\end{acknowledgments}

%%%%%%%%%%%%%%%%%%%%%%%%%%%%%%%%%%%%%%%%%%%%%%%%%%%%%%%%%%%%%%%%%%%%%%%%%%%%%%%%
\appendix

\section{Numerical approach}
\label{sec:FEM}

This Appendix describes the practical implementation of our numerical
approach for solving the spectral problems (\ref{eq:Psi_def}) and
(\ref{eq:PsiN_def}).

\subsection{Finite-element method}
\label{sec:FEM1}

\begin{figure}
\begin{center}
\includegraphics[width=50mm]{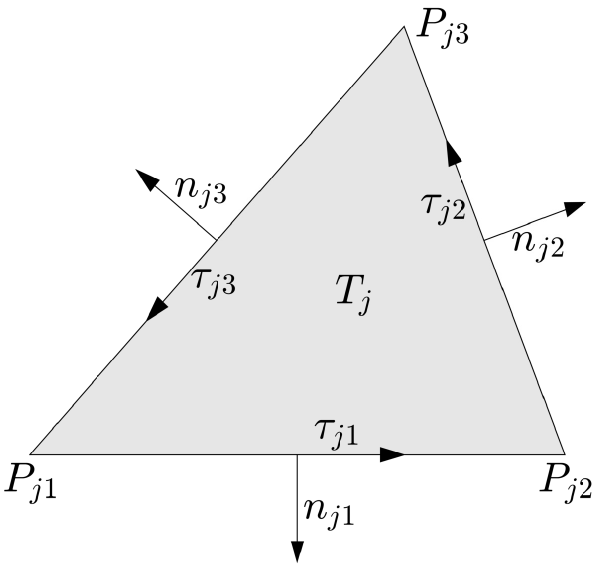} % {triangle.eps}
\end{center}
\caption{
Triangle $T_j$ of a mesh, its vertices $P_{jk}$ and normal and tangent
vectors $\n_{jk}$ and $\ttau_{jk}$.}
\label{fig:triangle}
% A_DN_patch_scheme;
\end{figure}

We employ a finite-element method to solve the integral eigenvalue
problems numerically.  The patch $\Gamma$ is covered by a triangular
mesh, composed of $N_t$ triangles denoted as $T_j$, $j =
1,\ldots,N_t$.  We denote three vertices of the $j$-th triangle as
$P_{jk} = (P_{jk}^x, P_{jk}^y)$, with $k=1,2,3$, and the total number
of vertices is $N_p$.  For convenience of notations, we allow the
index $k$ to take values $4$ and $5$ by setting $P_{j4} = P_{j1}$ and
$P_{j5} = P_{j2}$.  We also introduce the unit tangent vector
$\ttau_{jk}$ and the unit normal vector $\n_{jk}$ to the edge
$(P_{jk},P_{j(k+1)})$ (see Fig. \ref{fig:triangle}):
\begin{align}  \label{eq:tau_def}
\ttau_{jk} & = \frac{P_{j(k+1)} - P_{jk}}{|P_{j(k+1)} - P_{jk}|} \,, \\
\n_{jk} & = \pm (\ttau_{jk}^y, - \ttau_{jk}^x)^\dagger
\end{align}
(the sign $\pm$ is adjusted to ensure that $\n_{jk}$ is oriented
outward the triangle).

We employ linear basis functions to represent eigenfunctions on the
mesh in order to achieve higher accuracy as compared to
piecewise-constant basis functions.  For the sake of clarify, we start
with barycentric basis functions $\{\varphi_{jk}\}$, i.e.,
$\varphi_{jk}(\y)$ is a linear function on the $j$-th triangle $T_j$
(and $0$ otherwise) such that $\varphi_{jk}(P_{jk'}) = \delta_{k,k'}$
for $k,k' \in\{ 1,2,3\}$.  They form so-called discontinuous Galerkin
or ``broken P1'' space, to be distinguished from nodal basis (or hat)
functions that we will reconstruct at the end.  The barycentric basis
functions can be written explicitly as
\begin{equation}  \label{eq:varphi_def}
\varphi_{jk}(\y) = 1 - \langle \y-P_{jk}, \v_{jk}\rangle, 
\end{equation}
where $\langle \cdot, \cdot\rangle$ denotes the scalar product, and
\begin{equation}
\v_{jk} = \frac{1}{2|T_j|} \bigl(P^y_{j(k+2)} - P^y_{j(k+1)} , P^x_{j(k+2)} - P^x_{j(k+1)}\bigr)^\dagger
\end{equation}
ensures that $\varphi_{jk}(P_{j(k+1)}) = \varphi_{jk}(P_{j(k+2)}) = 0$
(with $|T_j|$ being the surface area of the $j$-th triangle).  One can
easily check that
\begin{equation}
\v_{jk} = \frac{|P_{j(k+2)} - P_{j(k+1)}|}{2|T_j|} \n_{j(k+1)} \,.
\end{equation}

We use the double index $jk$ to enumerate the basis functions, as well
as matrix elements constructed with the help of these functions.  An
eigenfunction $\Psi(\y)$ is searched as a linear superposition:
\begin{equation}
\Psi(\y) = \sum\limits_{j=1}^{N_t} \sum\limits_{k=1}^3 V_{jk} \varphi_{jk}(\y),
\end{equation}
with unknown coefficients $V_{jk}$.  Substituting this superposition
into the integral equation, multiplying it by another basis function
$\varphi_{j'k'}$ and integrating over $\Gamma$, we get the system of
$3N_t$ linear equations
\begin{equation}
\sum\limits_{j=1}^{N_t} \sum\limits_{k=1}^3 V_{jk} G_{jk,j'k'} = \frac{1}{\mu} \sum\limits_{j=1}^{N_t} \sum\limits_{k=1}^3 V_{jk} M_{jk,j'k'} ,
\end{equation}
where 
\begin{align} \nonumber
M_{jk,j'k'} & = \int\limits_{\Gamma} \varphi_{jk} \varphi_{j'k'} d\y = \delta_{j,j'} 
\int\limits_{T_j}  \varphi_{jk} \varphi_{j'k'} \, d\y  \\
& = \delta_{j,j'} \frac{|T_j|}{12} (1 + \delta_{k,k'})
\end{align}
is the standard mass matrix, and
\begin{equation}
G_{jk,j'k'} = \int\limits_{\Gamma} d\x \, \varphi_{j'k'}(\x) \int\limits_{\Gamma} \varphi_{jk}(\y) \, \G(\x,\y) d\y
\end{equation}
is the matrix representation of the kernel $\G(\x,\y)$ given by
Eq. (\ref{eq:G_kernel}).  

Up to now, this was a standard numerical procedure for an arbitrary
kernel $\G(\x,\y)$.  In the following, we rely on the explicit form of
the kernel and the flat geometry of the patch to obtain a
semi-analytical representation of the matrix elements $G_{jk,j'k'}$.
In fact, we aim to evaluate {\it exactly} the integral 
\begin{equation}
Q_{jk}(\x) = \int\limits_{\Gamma} \varphi_{jk}(\y)  \G(\x,\y) \, d\y.
\end{equation}
Using Eq. (\ref{eq:varphi_def}), we get
\begin{equation}
Q_{jk}(\x) = \frac{F_j(\x)}{2\pi} \bigl(1 + \langle P_{jk}, \v_{jk}\rangle \bigr) - \frac{H_j(\x,\v_{jk})}{2\pi} \,,
\end{equation}
where
\begin{align}  \label{eq:Fj_def}
F_j(\x) & = \int\limits_{T_j} \frac{d\y}{|\x-\y|}  \,,\\   \label{eq:Hj_def}
H_j(\x,\v) & = \int\limits_{T_j} \frac{\langle \v, \y \rangle \, d\y}{|\x-\y|} \,.
\end{align}
For the first integral, we use the identity $\Delta_{\y} |\x-\y| =
1/|\x-\y|$ to get by the divergence theorem
\begin{align} \nonumber
F_j(\x) & = \int\limits_{\partial T_j} \langle \n_{\y} , \nabla_{\y} |\x-\y|\rangle dl_{\y} \\
& = \sum\limits_{k=1}^3 \int\limits_{\partial T_{jk}} \frac{\langle \n_{\y} , (\y-\x)\rangle}{|\y-\x|} dl_{\y},
\end{align}
where $\partial T_{jk}$ denotes the edge of the $j$-th triangle
between vertices $P_{jk}$ and $P_{j(k+1)}$ (here $\Delta_{\y}$ is the
two-dimensional Laplace operator).  Using the parameterization $\y(t)
= P_{j,k} + (P_{j(k+1)}-P_{jk})t$ on this edge (with $t\in (0,1)$ and
$dl_{\y} = |\y'(t)| dt$), we have
\begin{equation}
F_j(\x) = \sum\limits_{k=1}^3 |P_{j(k+1)}-P_{jk}| \int\limits_0^1 \frac{\langle \n_{jk} , (\y(t)-\x)\rangle}{|\y(t)-\x|} dt,
\end{equation}
where $\n_{jk}$ is the outward unit normal vector to the edge
$(P_{jk}, P_{j(k+1)})$.  After lengthy but elementary computation of
the last integral, we get
\begin{equation}  \label{eq:Fj}
F_j(\x) = \sum\limits_{k=1}^3  \langle \n_{jk}, R_{jk}\rangle \, I_{jk} ,
\end{equation}
where $R_{jk} = P_{jk} - \x$,  and
\begin{equation}  \label{eq:Ijk}
I_{jk} = \ln \left( \frac{\langle R_{j(k+1)}, \ttau_{jk} \rangle + |R_{j(k+1)}|}{\langle R_{jk}, \ttau_{jk} \rangle + |R_{jk}|}\right),
\end{equation}
with the unit tangent vector $\ttau_{jk}$ given by
Eq. (\ref{eq:tau_def}).

For the integral (\ref{eq:Hj_def}), we use $\langle \v , \nabla_{\y}
|\x-\y|\rangle = \langle \v, (\y-\x) \rangle/|\y-\x|$ and the
divergence theorem to get
\begin{equation}
H_j(\x,\v) = \langle \v, \x \rangle F_j(\x) + \int\limits_{\partial T_j} |\x-\y| \langle \v, \n_{\y} \rangle \, d\y \,.
\end{equation}
As previously, we use the parameterization over each edge to get
\begin{equation}
H_j(\x,\v)  = \langle \v, \x \rangle F_j(\x) + \sum\limits_{k=1}^3 \langle \v, \n_{jk} \rangle J_{jk} ,
\end{equation}
where
\begin{align} \nonumber
J_{jk} & = \int\limits_{\partial T_{jk}} |\x-\y| \, dl_{\y} 
= \frac{I_{jk}}{2}  \bigl(|R_{jk}|^2 - \langle R_{jk},\ttau_{jk}\rangle^2\bigr) \\
& + \frac12 \biggl(\langle R_{j(k+1)},\ttau_{jk}\rangle |R_{j(k+1)}| - \langle R_{jk},\ttau_{jk}\rangle |R_{jk}|\biggr).
\end{align}

As the function $Q_{jk}(\x)$ is known exactly, the matrix elements can
be found as
\begin{equation}
G_{jk,j'k'} = \int\limits_{T_{j'}}  \varphi_{j'k'}(\x) \, Q_{jk}(\x) \, d\x.
\end{equation}
The simplest quadrature employs the barycentric rule.  However, one of
the major numerical difficulties for solving the spectral problem
(\ref{eq:Psi_integral}) is that the matrix $\GG$ is not sparse that
significantly limits its size and thus requires using relatively
coarse meshing.  In order to improve the accuracy, it is therefore
convenient to apply more accurate quadratures when integrating over
triangles.  We employ Dunavant 7-point quadrature rule
\cite{Dunavant85}: for a smooth enough function $f(\y)$, we set
\begin{equation}
\int\limits_{T_j} f(\y)\, d\y \approx |T_j| \sum\limits_{i=1}^7 w_i \, 
f\bigl(\nu_{i1} P_{j1} +  \nu_{i2} P_{j2} + \nu_{i3} P_{j3}\bigr),
\end{equation}
where 
\begin{align*}
w_1 & = 0.225, \quad \nu_{11} = \nu_{12} = \nu_{13} = 1/3 ,\\
w_2 & = 0.132394, ~~ \nu_{21} = 0.059716, ~~ \nu_{22} = \nu_{23} = 0.470142 , \\
w_5 & = 0.125939, ~~ \nu_{51} = 0.797427, ~~ \nu_{52} = \nu_{53} = 0.101287, 
\end{align*}
whereas $w_3 = w_4 = w_2$ with $\nu_{3i}$ and $\nu_{4i}$ being
obtained by permutations from $\nu_{2i}$, and $w_6 = w_7 = w_5$
with $\nu_{6i}$ and $\nu_{7i}$ being obtained by permutations
from $\nu_{5i}$.  This quadrature is known to be exact for
polynomials up to degree 5.

Once the matrix elements $G_{jk,j'k'}$ are found, one can solve
numerically the generalized eigenvalue problem for square matrices
$\GG$ and $\MM$, that allows one to approximate some Steklov
eigenvalues and eigenfunctions.  This final step can be further
improved by transforming the above representation in terms of
barycentric basis functions $\{\varphi_{jk}\}$, associated with
triangles, into more convenient nodal basis (or hat) functions
$\{\phi_i\}$, associated to the nodes of the mesh.  For each node
$P_i$ (with $i = 1,2,\ldots,N_p$), $\phi_i(\x)$ is a linear function
on the triangles that contain $P_i$ (and $0$ otherwise) such that
$\phi_i(P_i) = 1$ and $0$ at all other nodes.  In fact, on the $j$-th
triangle containing the vertex $P_i$, $\phi_i$ is simply the
barycentric function $\varphi_{jk}$, with the index $k$ such that
$P_{jk} = P_i$.  As a consequence, one can regroup the earlier defined
matrix elements $G_{jk,j'k'}$ and $M_{jk,j'k'}$ into new matrix
elements $\bar{G}_{i,i'}$ and $\bar{M}_{i,i'}$ as
\begin{align} \label{eq:Gbar}
\bar{G}_{i,i'} &= \sum\limits_{j,k: \atop P_{jk} = P_i} \sum\limits_{j',k': \atop P_{j'k'} = P_{i'}} G_{jk,j'k'}  \quad (i,i' = 1,\ldots,N_p), \\
\bar{M}_{i,i'} &= \sum\limits_{j,k: \atop P_{jk} = P_i} \sum\limits_{j',k': \atop P_{j'k'} = P_{i'}} M_{jk,j'k'}  \quad (i,i' = 1,\ldots,N_p).
\end{align}
In this way, one has to solve numerically the generalized eigenvalue
problem for square matrices $\bar{\GG}$ and $\bar{\MM}$: 
\begin{equation}
\bar{\GG} \VV = \lambda \bar{\MM} \VV .
\end{equation}
Let us denote the eigenvalues and eigenvectors of this problem as
$\lambda_k$ and $\VV_k$ and enumerate the eigenvalues to form a
decreasing sequence: $\lambda_0 \geq \lambda_1 \geq \lambda_2
\geq \cdots$.  One sees that $1/\lambda_k$ is an approximation of the
$k$-th eigenvalue $\mu_k$, whereas the components of the vector
$\VV_k$ are the coefficients of an approximate expansion of the $k$-th
eigenfunction $\Psi_k$ on the nodal basis functions:
\begin{equation}
\Psi_k(\y) = \sum\limits_{i=1}^{N_p} [\VV_k]_i \, \phi_i(\y).
\end{equation}
Moreover, the value of $\Psi_k$ on a node $P_i$ is simply $[\VV_k]_i$.

Note that the normalization of eigenfunctions is fixed by requiring
\begin{equation}
1 = \int\limits_{\Gamma} \Psi_k^2(\y) \, d\y = \sum\limits_{i,i'=1}^{N_p} [\VV_k]_i [\VV_k]_{i'} 
\underbrace{\int\limits_{\Gamma} \phi_i\, \phi_{i'}\, d\y}_{=\bar{M}_{i,i'}} .
\end{equation}
We also compute the projection of $\Psi_k$ onto a constant:
\begin{equation}
d_k = \int\limits_{\Gamma} \Psi_k(\y) \, d\y = \sum\limits_{i=1}^{N_p} [\VV_k]_i 
\underbrace{\int\limits_{\Gamma} \phi_i(\y)\, d\y}_{= \sum\limits_{j,k': P_{jk'} = P_i} |T_j|/3} ,
\end{equation}
from which the weights follow as $F_k = d_k^2/|\Gamma|$.  We outline
that the representation in terms of nodal basis functions has two
practical advantages: (i) the matrices $\bar{\GG}$ and $\bar{\MM}$ of
size $N_p \times N_p$ are much smaller than $\GG$ and $\MM$ (of size
$3N_t\times 3N_t$) that allows for their faster numerical
diagonalization; and (ii) the continuous nature of nodal basis
functions yields a smoother representation of eigenfunctions.  We
employ this method throughout the manuscript.

\subsection{Solving the alternative Steklov problem}

An equivalent reformulation of the spectral problem
(\ref{eq:PsiN_def}) was discussed in \cite{Henrici70,Grebenkov25b}.
To ensure that all eigenfunctions $\Psi_k^N$ with $k = 1,2,\ldots$ are
orthogonal to the constant function $\Psi_0^N = 1/\sqrt{|\Gamma|}$,
the kernel $\G(\y,\y')$ can be replaced by
\begin{equation}  \label{eq:GN_kernel}
\G^N(\y,\y') = \G(\y,\y') - \frac{\omega_\Gamma(\y) + \omega_\Gamma(\y')}{|\Gamma|} + \A_\Gamma,
\end{equation}
where $\omega_\Gamma(\y)$ and $\A_\Gamma$ were defined in
Eqs. (\ref{eq:w_def}, \ref{eq:A_def}).  In this way, the integral of
$\G^N(\y,\y')$ over $\y\in \Gamma$ vanishes so that a constant
function is an eigenfunction of the associated integral operator.  It
was shown in \cite{Grebenkov25b} that the eigenfunctions $\Psi_k^N$
satisfy for $k = 1,2,\ldots$:
\begin{equation}  \label{eq:PsiN_int}
\int\limits_{\Gamma} \G^N(\y,\y') \Psi_k^N(\y') d\y' = \frac{1}{\mu_k^N} \Psi_k^N(\y) \qquad (\y\in\Gamma).
\end{equation}
In other words, one can search for the eigenpairs $\{\mu_k^N,
\Psi_k^N\}$ (with $k \geq 1$) of the spectral problem
(\ref{eq:PsiN_def}) by solving the eigenvalue problem
(\ref{eq:PsiN_int}).  The missing eigenpair $\mu_0^N = 0$ and
$\psi_0^N = 1/\sqrt{|\Gamma|}$ can be added manually.

The numerical technique from Sec. \ref{sec:FEM1} is applicable for
solving the Steklov problem (\ref{eq:PsiN_def}), which is equivalently
described by the kernel $\G^N(\y,\y')$ from Eq. (\ref{eq:GN_kernel}).
In fact, we need to find
\begin{align} \nonumber
Q_{jk}^N(\x) & = \int\limits_{\Gamma} \varphi_{jk}(\y) \G^N(\x,\y) d\y  \\
& \approx Q_{jk}(\x) - \biggl[\frac{\omega_\Gamma(\x) + \omega_\Gamma(\x_j)}{|\Gamma|} - \A_\Gamma \biggr] \frac{|T_j|}{3} \,,
\end{align}
where $\x_j$ is the barycenter of $T_j$, and we used the barycentric
approximation $\int\nolimits_{T_j} \varphi_{jk}(\y) \,
\omega_\Gamma(\y) d\y \approx \omega_\Gamma(\x_j) \int\nolimits_{T_j} 
\varphi_{jk}(\y) \, d\y$ (it can further be improved via Dunavant 7-point
quadrature).  Using the approximations
\begin{equation}   \label{eq:wA_mesh}
\omega_{\Gamma}(\x) = \frac{1}{2\pi}\sum\limits_{j=1}^N F_j(\x)\,,  \quad 
\A_{\Gamma} \approx \frac{1}{|\Gamma|}\sum\limits_{j=1}^N |T_j| w(\x_j) ,
\end{equation}
we get
\begin{align} 
G_{jk,j'k'}^N & \approx 
\frac{|T_{j'}|}{3} Q_{jk}^N(\x_{j'}) \\ \nonumber
& \approx G_{jk,j'k'} - \frac{|T_j| \, |T_{j'}|}{9} \biggl[\frac{\omega_\Gamma(\x_j) + \omega_\Gamma(\x_{j'})}{|\Gamma|} - \A_\Gamma \biggr] \,,
\end{align}
from which the matrix elements $\bar{G}_{i,i'}^N$ are deduced via
Eq. (\ref{eq:Gbar}).  As previously, one searches for eigenpairs
$\{\lambda_k^N, \VV_k^N\}$ solving the generalized eigenvalue problem
$\bar{\GG}^N \VV_k^N = \lambda_k^N \bar{\MM} \VV_k^N$, enumerated by
$k = 1,2,\ldots$ (here the index starts from $1$, given that $\mu_0^N
= 0$).

Once the eigenfunction $\Psi_k^N(\y)$ is constructed on the patch, it
can be easily extended into the upper half-space $\R^3_+$.  For this
purpose, one can multiply Eq. (\ref{eq:PsiN_def}) by the Green's
function $G(\y,\y')$, multiply Eq. (\ref{eq:Green}) by
$\Psi_k^N(\y')$, subtract them, integrate over $\y' \in \R^3_+$, use
the Green's formula with boundary conditions and behavior at infinity
to get
\begin{equation}
\Psi_k^N(\y) = \Psi_k^N(\infty) + \mu_k^N \int\limits_{\Gamma} G(\y,\y') \Psi_k^N(\y') d\y,
\end{equation}
where $G(\y,\y')$ is given by the explicit formula
(\ref{eq:Green_half}).  In turn, the value at infinity,
$\Psi_k^N(\infty)$, can be determined by expressing $G(\y,\y')$ from
Eq. (\ref{eq:GN_kernel}) and using Eq. (\ref{eq:PsiN_int}):
\begin{equation}
\Psi_k^N(\infty) = - \frac{\mu_k^N}{|\Gamma|} \int\limits_{\Gamma} \omega_\Gamma(\y) \, \Psi_k^N(\y) d\y.
\end{equation}

\subsection{Validation}
\label{sec:validation}

To check the accuracy of our numerical method, we compute the Steklov
eigenvalues for the circular patch of unit radius.  This geometric
setting can be considered as the limit $c\to 0$ of an oblate spheroid
with semi-axes $1,1,c$, for which an efficient numerical method for
solving Steklov problems was developed in \cite{Grebenkov24}.  This
method employs the oblate spheroidal coordinates to achieve an
accurate matrix representation of the associated Dirichlet-to-Neumann
operator.  Its numerical diagonalization yields very accurate
eigenvalues, which will be referred to as ``exact''.

Table \ref{tab:validation} reports the first ten eigenvalues for both
versions of the Steklov problem (see also Fig. \ref{fig:disk_Vk} for
eigenfunctions $\Psi_k$).  Two different meshes were used to
illustrate the role of the mesh size on the accuracy.  The relative
error of the presented eigenvalues does not exceed $0.1\%$ even for
the coarser mesh.

The rotational symmetry of the patch implies that an eigenfunction on
$\Gamma$ can be represented in polar coordinates $(r,\phi)$ as
$e^{im\phi} v_{m,n}(r)$, with integer indices $m$ and $n$.  Moreover,
the structure of radial functions $v_{m,n}(r)$ is inherited from
spherical harmonics $Y_{m,n}$ (see \cite{Grebenkov24} for more
details); in particular, one can set $n = 0,1,2,\cdots$ and $m =
-n,-n+1,\cdots,n$, as shown in the bottom row of Table
\ref{tab:validation}.  Note that the eigenvalues with $(\pm m)n$
coincide.  As discussed in \cite{Grebenkov24}, the eigenfunctions
satisfying Neumann boundary condition (\ref{eq:Psi_Neumann})
correspond to even $m+n$.  Finally, as the difference $\G(\x,\y) -
\G^N(\x,\y)$ is a radial function (given that $\omega_\Gamma(\y)$ is a
function of $|\y|$), only axially symmetric eigenfunctions with $m =
0$ differ between the two versions of the Steklov problem.  This is
clearly seen from Table \ref{tab:validation}.  Note also that our
approach does not recover the eigenvalue $\mu_0^N = 0$ that
corresponds to a constant eigenfunction.

\begin{table*}
\centering
\begin{tabular}{|c|c|c|c|c|c|c|c|c|c|c|c|c|} \hline
$N_p$ & $N_t$ & $k$  &  0       &   1      &   2      &   3      &    4     &   5     &   6      &   7      &   8      &   9      \\  \hline       
375 &  688    &  D   &  1.1588  &  2.7573  &  2.7573  &  4.1252  &  4.1252  &  4.3209 &  5.4053  &  5.4053  &  5.8984  &  5.8984  \\
553 & 1032    &  D   &  1.1585  &  2.7565  &  2.7565  &  4.1240  &  4.1240  &  4.3196 &  5.4036  &  5.4036  &  5.8963  &  5.8963  \\
    & exact   &  D   &  1.1578  &  2.7548  &  2.7548  &  4.1214  &  4.1214  &  4.3169 &  5.4003  &  5.4003  &  5.8924  &  5.8924  \\ \hline \hline
375 & 688     &  N   &          &  2.7573  &  2.7573  &  4.1252  &  4.1252  &{\clb 4.1239}&  5.4053  &  5.4053  &  5.8984  &  5.8984  \\
553 & 1032    &  N   &          &  2.7565  &  2.7565  &  4.1240  &  4.1240  &{\clb 4.1230}&  5.4036  &  5.4036  &  5.8962  &  5.8963  \\
    & exact   &  N   &  0       &  2.7573  &  2.7573  &  4.1252  &  4.1252  &{\clb 4.1213}&  5.4003  &  5.4003  &  5.8924  &  5.8924  \\  \hline
    &         & $mn$ &  $00$    &  $(-1)1$ &   $11$   &  $(-2)2$ &   $22$   &   $02$  &  $(-3)3$ &  $33$    &  $(-1)3$ &  $13$   \\  \hline       
\end{tabular}
\caption{
Comparison of the Steklov eigenvalues for the circular patch of unit
radius between our FEM method and semi-analytical computation via
oblate spheroidal coordinates in \cite{Grebenkov24,Grebenkov25b},
referred to as ``exact''.  The number of vertices ($N_p$) and the
number of triangle ($N_t$) are indicated, whereas the letters D and N
distinguish respectively the spectral problems (\ref{eq:Psi_def}) and
(\ref{eq:PsiN_def}).  The bottom line indicates an alternative
enumeration of the eigenpairs according to the symmetries of the
eigenfunctions.  For the ``Neumann version'', the eigenvalue $4.1213$
and its numerical estimates were placed into the column with $k = 5$
to respect the radial symmetry $mn = 02$ of the associated
eigenfunction, despite the artificially induced error in ordering of
the eigenvalues.}
\label{tab:validation}
% [Ed,V,dk, p,e,t,Gam] = A_DN_patch_FEM_ellipses(1);
\end{table*}
%\end{widetext}

\section{Computation of the constant $\A_\Gamma$}
\label{sec:wGamma}

The function $\omega_\Gamma(\x)$ from Eq. (\ref{eq:w_def}) and the
constant $\A_\Gamma$ from Eq. (\ref{eq:A_def}) play an important role
in the reformulation of the spectral problem (\ref{eq:PsiN_def}); see
also \cite{Grebenkov25b} for various applications of these quantities.
For a circular patch, their exact forms were obtained in
\cite{Grebenkov25b}:
\begin{equation}  \label{eq:AGamma_disk}
\omega_\Gamma(\x) = \frac{2}{\pi} E(|\x|) , \qquad \A_\Gamma = \frac{8}{3\pi^2} \,,
\end{equation}  %{\clb [Confirmed, see \verb|[w,A] = A_DN_patch_w_check_disk();|]}
where $E(z)$ is the complete elliptic integral of the second kind.  In
this Appendix, we discuss an alternative computation of these
quantities for other domains, beyond the basic FEM representation
(\ref{eq:wA_mesh}).

\subsection{General patch}

For any triangular patch, the function $\omega_\Gamma(\x)$ is simply
proportional to the function $F_j(\x)$ defined by
Eq. (\ref{eq:Fj_def}), for which the representation (\ref{eq:Fj})
provides the exact solution.  Moreover, the exact representation
(\ref{eq:Fj}) can be generalized to any polygon.  As a consequence,
this representation allows one to compute {\it exactly} the function
$w_{\Gamma}(\x)$ from Eq. (\ref{eq:w_def}) for any polygonal patch,
without resorting to triangulations:
\begin{equation}  \label{eq:omega_polygon}
\omega_\Gamma(\x) = \frac{1}{2\pi} \sum\limits_k \langle \n_k, (P_k - \x)\rangle I_k ,
\end{equation}
where $P_k$ are the vertices of the polygon, $\n_k$ is the unit normal
vector to the $k$-th edge between vertices $P_k$ and $P_{k+1}$ ($\n_k$
is oriented outward the polygon), $I_k$ is given by
Eq. (\ref{eq:Ijk}), and the sum is over all edges.  In turn, the
constant $\A_\Gamma$ can be written as follows
\begin{align*}
\A_\Gamma & = \frac{1}{2\pi |\Gamma|^2} \int\limits_{\partial\Gamma} dl_{\y} \langle \n_{\y}, \nabla_{\y} \rangle
\int\limits_{\Gamma} |\x-\y|\, d\x  \\
& = \frac{1}{2\pi |\Gamma|^2} \int\limits_{\partial\Gamma} dl_{\y} \langle \n_{\y}, \nabla_{\y} \rangle
\int\limits_{\partial \Gamma}  \langle \n_{\x} , \nabla_{\x} |\x-\y|^3/9 \rangle dl_{\x} ,
\end{align*}
where we used that $\Delta_{\x} |\x-\y|^3/9 = |\x-\y|$.  As a
consequence, we obtain
\begin{align} \nonumber
\A_\Gamma & = \frac{1}{6\pi |\Gamma|^2} \int\limits_{\partial\Gamma} dl_{\y} \langle \n_{\y}, \nabla_{\y} \rangle
\int\limits_{\partial \Gamma}  \langle \n_{\x} , \nabla_{\x} |\x-\y|(\x-\y)\rangle dl_{\x} \\ \nonumber
& = - \frac{1}{6\pi |\Gamma|^2} \int\limits_{\partial\Gamma \times \partial\Gamma}  
\biggl(\langle \n_{\x}, \n_{\y} \rangle |\x-\y| \\  \label{eq:AGamma_general}
& + \frac{\langle \n_{\x}, \x-\y\rangle \, \langle \n_{\y}, \x-\y\rangle}{|\x-\y|} \biggr) dl_{\x}  dl_{\y}.
\end{align}
This formula opens a purely geometric way to access $\A_\Gamma$,
allowing one to skip the computation of $\omega_\Gamma(\x)$ and its
integration.
We recall that the constant $\A_\Gamma$ determines via
Eq. (\ref{eq:A_def}) the coefficient $c_2$ of the Taylor expansion
(\ref{eq:Cmu_Taylor}) of the reactive capacitance. 
% see A_DN_patch_A_rectangle(a,b);

\subsection{Rectangular patch}

For a rectangle $(-a,a)\times (-b,b)$, Eq. (\ref{eq:omega_polygon}) yields
\begin{align}
& \omega_\Gamma(x,y) = \frac{1}{2\pi} \\  \nonumber
& \times \biggl\{
   (b+y)\ln \biggl(\frac{a-x+ \sqrt{(a-x)^2+(b+y)^2}}{-a-x+\sqrt{(a+x)^2+(b+y)^2}}\biggr)  \\  \nonumber
& +(a-x)\ln \biggl(\frac{b-y+ \sqrt{(a-x)^2+(b-y)^2}}{-b-y+\sqrt{(a-x)^2+(b+y)^2}}\biggr)  \\  \nonumber
& +(b-y)\ln \biggl(\frac{a+x+ \sqrt{(a+x)^2+(b-y)^2}}{-a+x+\sqrt{(a-x)^2+(b-y)^2}}\biggr)  \\  \nonumber
& +(a+x)\ln \biggl(\frac{b+y+ \sqrt{(a+x)^2+(b+y)^2}}{-b+y+\sqrt{(a+x)^2+(b-y)^2}}\biggr)\biggr\}  .
\end{align}  %{\clb See \verb|[w,pet,A] = A_DN_patch_w_rectangle_show(1,0.1);|}
After a lengthy but elementary integration of this expression over the
rectangle $(-a,a)\times (-b,b)$, we deduce the associated constant
$\A_\Gamma$:
\begin{equation}
\A_\Gamma = \frac{a \, \eta(b/a)}{\pi b^2} + \frac{b \, \eta(a/b)}{\pi a^2} \,, 
\end{equation}
where  
\begin{align}  \nonumber
\eta(y) & = \frac{1 + y^3 - (1+y^2)^{3/2}}{3} + \frac{\sqrt{1+y^2}-1}{2} \\
& + \frac{y^2}{4}\ln \biggl(\frac{1 + \sqrt{1+y^2}}{-1 + \sqrt{1+y^2}}\biggr).
\end{align}
For instance, we get for the square of edge $2b$
\begin{equation}  \label{eq:AGamma_square}
\A_\Gamma = \frac{2\eta(1)}{\pi b} \approx \frac{0.2366}{b} \,, 
\end{equation}
with $\eta(1) = \tfrac{1}{6} (1-\sqrt{2}) + \tfrac12 \ln(1 + \sqrt{2})
\approx 0.3717$.

In the limit $a\to 0$, we use the asymptotic behavior of the function
$\eta(y)$, namely, $\eta(y) \simeq y/2 + O(1)$ as $y\to \infty$ and
$\eta(y) \simeq y^2(-1 + 2\ln(2/y))/4 + O(y^3)$ as $y\to 0$, to obtain
\begin{equation}  \label{eq:A_rectangle_asympt}
\A_\Gamma \simeq \frac{1}{\pi b} \biggl(\frac{1+2\ln 2}{4} + \frac12 \ln (b/a)\biggr) + O(a/b).
\end{equation}
In fact, as the rectangular patch shrinks to an interval, which is
inaccessible to reflected Brownian motion, the constant $\A_\Gamma$
diverges logarithmically.

\subsection{Elliptic patch}

For an elliptic patch with semi-axes $a$ and $b$, we employ the
parameterization $\x_1(t_1) = (a \cos t_1, b \sin t_1)^\dagger$ so that
$c(t_1) = |\tfrac{d}{dt_1} \x_1(t_1)| = \sqrt{a^2 \sin^2 t_1 + b^2
\cos^2 t_1}$ and $\n_{\x_1} = (b \cos t_1, a\sin t_1)^\dagger/c(t_1)$ and
similar for $\x_2(t_2)$.  As a consequence,
Eq. (\ref{eq:AGamma_general}) reads
\begin{align} \nonumber
& \A_\Gamma = \frac{-1}{6\pi^3 b} \int\limits_0^{2\pi} dt_1 \int\limits_0^{2\pi} dt_2
\biggl\{ (\gamma^{-2} \cos t_1 \cos t_2 + \sin t_1 \sin t_2) \\  \label{eq:A_ellipse}
& \times D(t_1,t_2) - \frac{[1 - (\cos t_1 \cos t_2 + \sin t_1 \sin t_2)]^2}{D(t_1,t_2)} \biggr\} ,
\end{align}
where $\gamma = a/b$ and
\begin{equation*}
D(t_1,t_2) = \sqrt{\gamma^2 (\cos t_1 - \cos t_2)^2 + (\sin t_1 - \sin t_2)^2} \,.
\end{equation*}

To proceed, one can split the integration domain
$(0,2\pi)\times(0,2\pi)$ into four twice smaller squares and use the
symmetry of sine and cosine functions to reduce the integration domain
to $(0,\pi)\times (0,\pi)$.  By introducing the new integration
variables $t_\pm = (t_1 \pm t_2)/2$, these integrals can be reduced to
the complete elliptic functions.  Skipping these tedious lengthy
computations, we just provide the final result, which takes a
particularly simple form:  
% see the file 'ellipse_A.tex' for details
%
\begin{equation}  \label{eq:AGamma_ellipse}
\A_\Gamma = \frac{16}{3\pi^3 b} K\bigl(\sqrt{1-a^2/b^2}\bigr) ,
\end{equation}
where $K(z)$ is defined by Eq. (\ref{eq:ellipticK}).  In the
particular case of a circular patch ($a = b$), we use $K(0) = \pi/2$
to retrieve the result (\ref{eq:AGamma_disk}).  In turn, in the limit
$a\to 0$, we use the asymptotic relation (\ref{eq:K_asympt}) to get
\begin{equation}
\A_\Gamma \simeq \frac{16}{3\pi^3 b} \ln(4b/a)  \quad (a\ll b).
\end{equation}

Substituting Eq. (\ref{eq:AGamma_ellipse}) into our upper bound
(\ref{eq:mu0_boundsA}), we get for any $0 < a \leq b$:
\begin{equation}  \label{eq:mu0_ellipse_bound1}
\mu_0 \leq \frac{3\pi^2}{16 a K(\sqrt{1-a^2/b^2})} \,.
\end{equation}
It is instructive to compare this inequality with the Payne's upper
bound (\ref{eq:mu0_upper}).  Substituting Eq. (\ref{eq:Cinf_ellipse})
for the electrostatic capacitance, one has
\begin{equation}  \label{eq:mu0_ellipse_bound2}
\mu_0 \leq \frac{2}{a K(\sqrt{1-a^2/b^2})} \,.
\end{equation}
Remarkably, both bounds have the same functional form and differ only
by the numerical prefactors: $3\pi^2/16 \approx 1.85$ in
Eq. (\ref{eq:mu0_ellipse_bound1}) and $2$ in
Eq. (\ref{eq:mu0_ellipse_bound2}).  One sees that our upper bound is
slightly more accurate for any $0 < a \leq b$.

\subsection{Circular annulus}

Finally, we consider a circular annulus of radii $R_1 < R_2$.  As the
boundary of the annulus is split into two circles, the integral in
Eq. (\ref{eq:AGamma_general}) has four contributions when $\x_1$ and
$\x_2$ run over the inner/outer circles:
\begin{equation*}
\A_{\Gamma} = \A^{ii}_{\Gamma} + \A^{oo}_{\Gamma} + \A^{io}_{\Gamma} + \A^{oi}_{\Gamma} ,
\end{equation*}
where superscripts indicate inner and outer circles.  According to
Eq. (\ref{eq:AGamma_disk}), we get
\begin{equation}
\A^{ii}_{\Gamma} + \A^{oo}_{\Gamma} = \frac{8}{3} \, \frac{R_1^3 + R_2^3}{|\Gamma|^2} \,,
\end{equation}
with $|\Gamma| = \pi (R_2^2 - R_1^2)$.  We now compute the remaining
contributions.

For the outer circle, we employ the parameterization $\n_{\x_1} =
(\cos t_1, \sin t_1)^\dagger$, $\x_1(t_1) = R_2 \n_{\x_1}$, $dl_{\x_1}
= R_2 dt_1$, and similar for $\x_2$.  For the inner circle, we have
$\n_{\x_1} = -(\cos t_1, \sin t_1)^\dagger$, $\x_1(t_1) = - R_1
\n_{\x_1}$, $dl_{\x_1} = R_1 dt_1$, and similar for $\x_2$.  We get
then
\begin{align*}
\A^{io}_{\Gamma} & = \frac{R_1 R_2 \sqrt{R_1^2+R_2^2}}{3|\Gamma|^2} \int\limits_0^{2\pi}  
\biggl(\cos(t) \sqrt{1 - \gamma \cos(t)} \\
& + \frac{\cos(t) - \tfrac{\gamma}{2} (1+\cos^2(t))}{\sqrt{1 - \gamma \cos(t)}}\biggr) dt,
\end{align*}
where $\gamma = 2R_1 R_2/(R_1^2 + R_2^2) < 1$.  The last integral can
be expressed in terms of the complete elliptic integrals as
\begin{align} \nonumber
\A^{io}_{\Gamma} & = \frac{4R_1 R_2 \sqrt{R_1^2+R_2^2}}{3|\Gamma|^2 \gamma\sqrt{1+\gamma}} 
 \biggl((1-\gamma^2)K(\sqrt{2\gamma/(1+\gamma)}) \\
& - (1+\gamma)E(\sqrt{2\gamma/(1+\gamma)})\biggr).
\end{align}  % see  A_DN_patch_A_annulus_check();
Using the Landen's transformations, one gets
\begin{align*}
& K( \sqrt{2\gamma/(1+\gamma)} ) = \frac{2 K((1 - \sqrt{1-\gamma^2})/\gamma)}{1+ \sqrt{(1-\gamma)/(1+\gamma)}}  \\
& \quad = (1 + R_1/R_2) K( R_1/R_2 ) ,\\
& E( \sqrt{2\gamma/(1+\gamma)} ) = \biggl(1 + \sqrt{\frac{1-\gamma}{1+\gamma}}\biggr) E((1 - \sqrt{1-\gamma^2})/\gamma) \\
& \quad - \frac{2}{1 + \sqrt{\frac{1+\gamma}{1-\gamma}}} K((1 - \sqrt{1-\gamma^2})/\gamma) \\
& \quad = \frac{2E(R_1/R_2)}{1+R_1/R_2} - (1 - R_1/R_2) K(R_1/R_2).
\end{align*}
Substituting these representations, we get after simplifications:
\begin{equation*} 
\A^{io}_{\Gamma} = \frac{4R_2 \bigl[ (R_2^2-R_1^2) K( R_1/R_2 ) - (R_2^2+R_1^2) E( R_1/R_2 ) \bigr]}{3|\Gamma|^2} .
%\biggl( (R_2^2-R_1^2) K( R_1/R_2 ) - (R_2^2+R_1^2) E( R_1/R_2 ) \biggr).
\end{equation*}  % see  A_DN_patch_A_annulus_check();
By symmetry, one also gets $\A^{oi}_{\Gamma} = \A^{io}_{\Gamma}$ that
completes the computation of $\A_\Gamma$:
\begin{align}  \nonumber
\A_\Gamma & = \frac{8}{3\pi^2 (R_2^2-R_1^2)^2} \biggl( R_1^3 + R_2^3 + R_2 (R_2^2-R_1^2) K( R_1/R_2 ) \\
& - R_2 (R_2^2+R_1^2) E( R_1/R_2 ) \biggr) .
\end{align}

In contrast, there is no an explicit exact representation of the
electrostatic capacitance $C(\infty)$ of a circular annulus (see the
discussion in Sec. 8.6 of \cite{Sneddon}); an approximate formula was
given in \cite{Smythe51}. 

For a thin annulus, one can set $R_1 = R$ and $R_2 = R + a$.  Using
the asymptotic behavior of the complete elliptic integrals $K(z)$ and
$E(z)$ as $z\to 1$, we get after simplifications the small-$a$
behavior of $\A_\Gamma$:
\begin{equation}
\A_\Gamma \approx \frac{\ln(8R/a) + 3/2}{2\pi^2 R} \,.
\end{equation}  % see 'annulus.mw'
% see also [A] = A_DN_patch_A_annulus_fig;
Comparing this expression with Eq. (\ref{eq:A_rectangle_asympt}) for a
thin rectangle, we conjecture the following leading-order dependence
for any thin patch, which can be seen as a ``thickening'' of a planar
curve $C$ by ``width'' $2a$ (i.e., $\Gamma = \{\x\in \partial \R^3_+
~:~ |\x - C| < a\}$):
\begin{equation}
\A_\Gamma \simeq \frac{2\ln(1/a)}{\pi |\partial \Gamma|} + O(1) \quad (a\to 0),
\end{equation}
where $|\partial\Gamma|$ is the perimeter of the patch.  Combining
this relation with the bounds (\ref{eq:mu0_boundsA}) and assuming that
$F_0$ is close to $1$, we get another conjecture that
\begin{equation}
\frac{1}{a \mu_0} \simeq \frac{2}{\pi} \ln (1/a) + O(1)  \,,
\end{equation}
independently of the actual shape and length of the thin patch, which
affect the constant (subleading) term.
%\verb|[Ed,V,dk,Fk, p,e,t,Gam,Geom] = A_DN_patch_FEM_annulus(0.9,1, 0, 20);|
%]}

\end{document}